\documentclass[acmcsur]{x}
\usepackage{graphicx}

\newcommand{\ReL}{\hbox{$\mathcal R\!\raise2pt\hbox{$\varepsilon$}\!\hbox{$\mathcal L$}$}}
\newcommand{\ReLinda}{\hbox{$\mathcal R\!\raise2pt\hbox{$\varepsilon$}\!\hbox{$\mathcal L$inda}$}}
\newcommand{\SA}{\hbox{\sc sa}}
\newcommand{\SC}{\hbox{\sc sc}}
\newcommand{\Aty}{\hbox{\sc a}}
\usepackage[dvips]{epsfig}
\def\FT{Fault-\kern-2pt Tolerant}
\def\WWW{World-\kern-1pt Wide Web}

\def\nmr{\hbox{$N$\kern-1pt MR}}
\def\ImAlive{$<$I'm Alive$>$}

\font\sevenrm=cmr7
\font\nineit=cmti9
\let\mc=\ninerm 
\newbox\PPbox 
\setbox\PPbox=\hbox{\kern.5pt\raise1pt\hbox{\sevenrm+\kern-1pt+}\kern.5pt}
\def\PP{\copy\PPbox}
\def\CPLUSPLUS{{\mc C\kern-1pt\PP\spacefactor1000}}
\def\ICPLUSPLUS{{\nineit C\kern-1pt\mc\PP\spacefactor1000}}
\newlength{\widthplusfourty}
\newcommand{\ths}[3]{\hbox{$#1$T/$#2$H/$#3$S}}

\newcommand{\CPP}{$\hbox{C}\hskip-1pt\hbox{+}\hskip-1pt\hbox{+}$}

\begin{document}
\title{A Survey of Linguistic Structures for Application-level Fault-Tolerance}
\author{VINCENZO DE FLORIO and CHRIS BLONDIA\\
University of Antwerp\\
Department of Mathematics and Computer Science\\
Performance Analysis of Telecommunication Systems group\\
Middelheimlaan 1, 2020 Antwerp, Belgium, \emph{and}\\
Interdisciplinary institute for BroadBand Technology\\
Gaston Crommenlaan 8, 9050 Ghent-Ledeberg, Belgium}
\category{D.2.7}{Software Engineering}%
                {Software Architectures}[Languages \and
                                         Domain-specific architectures]
\category{D.3.2}{Programming Languages}%
                {Language Classifications}[Specialized application languages]
\category{K.6.3}{Management of Computing and Information Systems}%
                {Software Management}[Software development \and
                                      Software maintenance]
\category{D.2.2}{Software Engineering}%
                {Design Tools and Techniques}[Software libraries]
\category{D.2.7}{Software Engineering}%
                {Distribution, Maintenance, and Enhancement}[Portability]
%
\begin{abstract}
The structures for the expression of fault-tolerance
provisions into the application software are
the central topic of this paper.
Structuring techniques answer the questions ``How to
incorporate fault-tolerance in the application layer
of a computer program''
and ``How to manage the fault-tolerant code''. As such, they
provide means to control complexity,
the latter being a relevant factor for the introduction
of design faults.
This fact and the ever increasing
complexity of today's distributed software justify
the need for simple, coherent, and effective
structures for the expression of fault-tolerance in the application
software. In this text we first
define a ``base'' of structural attributes
with which application-level fault-tolerance structures
can be qualitatively assessed and compared with each other
and with respect to the above mentioned needs.
This result is then used to provide an elaborated
survey of the state-of-the-art of application-level
fault-tolerance structures.
\end{abstract}

\maketitle

\section{Introduction}\label{Chap:Intro}
Research in fault-tolerance has focused for decades on \emph{hardware\/}
fault-tolerance, i.e., on devising a number of effective and ingenious hardware
solutions to cope with physical faults. 
For several years this approach was considered as the only one effective solution
to reach the required availability and data integrity demanded
by ever more complex computer services. We now know that this is not true.
Hardware fault-tolerance is an important requirement to tackle the problem,
but cannot be considered as the only way to go:
Adequate tolerance of faults and end-to-end fulfilment
of the dependability design goals of a complex software system
must include means to avoid, remove, or tolerate faults that
operate at all levels, \emph{including the application layer}.

While effective solutions have been found, e.g., for the hardware~\cite{Pra96},
the operating system~\cite{Den76}, or the middleware~\cite{omg98} layers,
the problem of an effective system structure to express fault-tolerance
provisions in the application layer of computer programs is
still an open one. To the best of our knowledge, no detailed
critical survey of the available solutions exists. Through this paper
the authors attempt to fill in this gap. Our target topic 
is linguistic structures for application-level fault-tolerance, so we
do not tackle herein other important approaches that do not
operate at the application-level, such as the fault-tolerance models
based on transparent task replication~\cite{GuSc97},
as e.g. in CORBA-FT~\cite{omg98}. Likewise this text does not
include approaches such as the one in~\cite{EbKi04}, where the focus is on
automating the transformation of
a given fault-intolerant program into a fault-tolerant
program. The reason behind this choice is that, due to their
exponential complexity, those methods are currently only effective
when the state space of the target program is very 
limited~\cite{MSU-CSE-00-13}.

Another important goal of this text is to pinpoint the consequences of
inefficient solutions to the aforementioned problem as well as to
increase the awareness of a need for an optimal solution to it:
Indeed, the current lack of a simple and coherent system structure 
for software fault-tolerance engineering able to provide the designer
with effective support towards fulfilling goals such as
maintainability, re-usability, and service portability
of fault-tolerant software manifests itself as a true bottleneck
for system development.

The structure of this paper is as follows: in Sect.~\ref{s:rat} we
explain the reasons behind the need for system-level fault-tolerance.
There we also provide a set of key desirable attributes for
a hypothetically perfect linguistic structure for application-level
fault-tolerance (ALFT).
Section~\ref{Chap:DesignTools} is a detailed survey of modern available
solutions, each of which is qualitatively assessed with respect to
our set of attributes. Some personal considerations and conjectures
on what is missing and possible ways to achieve it
are also part of this section.
Section~\ref{Chap:end} finally summarizes 
our conclusions and provides a comparison of the reviewed approaches.

\section{Rationale}\label{s:rat}
If in the early days of modern computing it was to some extent acceptable
that outages and wrong results occurred rather often\footnote{This excerpt from
        a report on the ENIAC activity~\cite{Weik61} 
        gives an idea of how dependable computers were in 1947:
        ``power line fluctuations and power failures made
        continuous operation directly off transformer mains an impossibility [\ldots]
        down times were long; error-free running periods were short [\ldots]''.
        After many considerable improvements, still
        ``trouble-free operating time remained at about 100 hours a week during 
        the last 6 years of the ENIAC's use'', i.e., an availability of about 60\%!},
being the main role of computers basically that of a fast solver of numerical problems,
today the criticality associated with many tasks dependent on computers
requires strong guarantees for properties such as availability and data integrity.

A consequence of this growth in complexity and criticality is the need for
\begin{itemize}
\item techniques to assess and enhance, in a 
justifiable way, the reliance to be placed on the services provided by computer systems, and
\item techniques to lessen the risks associated with
computer failures, or at least bound the extent of
their consequences. 
\end{itemize}
This paper focuses in particular on
\textbf{fault-tolerance}, that is, how to ensure a service up to fulfilling
the system's function even in the presence of ``faults''~\cite{ALRL04,ALR04}.
A fault is a defect, or an imperfection, or a lack in a system's 
hardware or software component. It is generically defined as the adjudged
or hypothesised cause of an error. Faults can have their origin within the
system boundaries (internal faults) or outside, i.e., in the
environment (external faults). In particular, an internal fault
is said to be active when it produces an error,
and dormant (or latent) when it does not.
A dormant fault becomes an active fault when it is activated by either its
process or the environment. Fault latency is defined as either the
length of time between the 
occurrence of a fault and the appearance of the corresponding error, or
the length of time between the occurrence of a fault and its removal.

Faults can be classified according to so-called 
viewpoints~\cite{Lapr92,Lapr95,Lapr98}---phenome\-no\-lo\-gical
cause, nature, phase of creation or occurrence, situation with respect to system
boundaries, persistence. Not all combinations can give rise to a fault class---this
process only defines 17 {fault classes}, summarised in
Fig.~\ref{Fig:FaultClasses}.
These classes have been further partitioned into three groups, known as 
combined fault classes:

\begin{figure}
\centerline{\includegraphics[width=14cm]{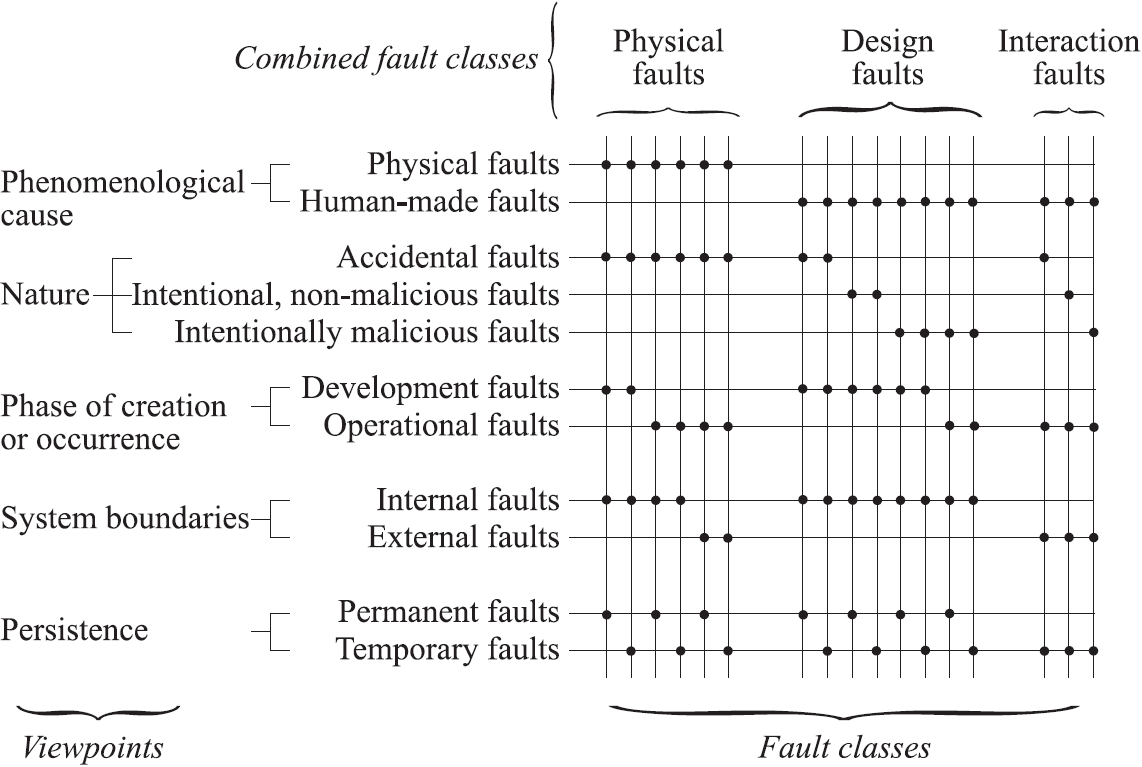}}
\caption{Laprie's fault classification scheme.}
\label{Fig:FaultClasses}
\end{figure}

\begin{description}
  \item[Physical faults:] \ 

  \begin{itemize}
  \item Permanent, internal, physical faults. This class concerns those faults that have their
  origin within hardware components and are continuously active. A typical example is given
  by the fault corresponding to a worn out component.
  \item Temporary, internal, physical faults (also known as
  {intermittent faults})~\cite{Bon97}.
  These are typically internal, physical defects that become active depending on a
  particular pointwise condition. 
  \item Permanent, external, physical faults. These are faults induced on the system by the
  physical environment. 
  \item Temporary, external, physical faults (also known as {transient faults})~\cite{Bon97}.
  These are faults induced by environmental phenomena, e.g., EMI.
  \end{itemize}
  \item[Design faults:] \ 

  \begin{itemize}
  \item Intentional, though not malicious, permanent / temporary design faults. These are
  basically trade-offs introduced at application-layer design time. A typical example is
  insufficient dimensioning
  (underestimations of the size of a given field in a communication
  protocol\footnote{A noteworthy example is given by the bad dimensioning
	  of IP addresses, which gave raise to IPv6.},
  and so forth).
  \item Accidental, permanent, design faults (also called 
  systematic faults, or Bohrbugs): flawed algorithms 
  that systematically turn into the same errors in the presence of the
  same input conditions and initial states---for instance,
  an unchecked divisor that can result in a division-by-zero error.
  \item Accidental, temporary design faults (known as 
  Heisenbugs\label{Pag:Heisenbugs}, for ``bugs of Heisenberg'', 
  after their elusive character): while systematic faults have an evident, 
  deterministic behaviour, these bugs depend on subtle combinations of the system
  state and environment. 
  \end{itemize}
  \item[Interaction faults:] \ 
  \begin{itemize}
  \item Temporary, external, operational, human-made, accidental faults.
  These include operator faults, in which an operator does not
  correctly perform his or her role in system operation.
  \item Temporary, external, operational, human-made, non-malicious faults:
  ``neglect, interaction, or incorrect use problems''~\cite{Sib98}.
  Examples include poorly chosen passwords and bad system parameter setting.
  \item Temporary, external, operational, human-made, malicious faults.
  This class includes the so-called malicious replication faults, i.e., faults
  that occur when replicated
  information in a system becomes inconsistent, e.g. because
  the processes that are supposed to provide identical results no longer do so.
  \end{itemize}
\end{description}

\subsection{A Need for Software Fault-Tolerance}\label{anfsft}
Research in fault-tolerance concentrated for many years on hardware
fault-tolerance, i.e., on devising a number of effective and ingenious hardware
structures to cope with faults~\cite{John89a}. 
For some time this approach was considered as the only one needed in order
to reach the requirements of availability and data integrity demanded
by modern complex
computer services. Probably the first researcher who realized that this was
far from being true was B. Randell who in~\cite{Ran75} questioned hardware 
fault-tolerance as the only approach to pursue---in the cited paper he states: 

\begin{quote}\label{Quote:Randell}
``Hardware component failures are only \emph{one\/} source of unreliability in computing
  systems, decreasing in significance as component reliability improves, while software faults
  have become increasingly prevalent with the steadily increasing size and complexity of
  software systems.''
\end{quote}

Indeed most of the complexity supplied by modern computing services lies in their
software rather than in the hardware layer~\cite{Lyu98a,Lyu98b,HuKi95,Wien93,Ran75}.
This state of things could only be reached
by exploiting a powerful conceptual tool
for managing complexity in a flexible and effective way, i.e., 
devising hierarchies of sophisticated abstract machines~\cite{Tan90}.
This translates into implementing software with high-level computer languages lying
on top of other software strata---the device drivers layers, the basic services kernel,
the operating system, the run-time support of the involved programming languages,
and so forth. 

Partitioning the complexity into stacks of software layers allowed the implementor
to focus exclusively on the high-level aspects of their problems, and hence
it allowed them to manage greater and greater degrees of complexity. But though made transparent,
this complexity is still part of the overall system being developed. 
A number of complex 
algorithms are executed by the hardware at the same time, resulting in the simultaneous 
progress of many system states---under the hypothesis that no involved abstract machine 
nor the actual hardware be affected by faults. Unfortunately, as in real life 
faults do occur, the corresponding deviations are likely to jeopardise 
the system's function, also propagating from one layer to the other,
unless appropriate means are taken to avoid in the first place, or
to remove, or to tolerate these faults.
In particular, faults may also occur in the
\textbf{application layer}, that is, in the abstract machine on top of the
software hierarchy\footnote{In what follows, the application layer is to be
    intended as the programming and execution context in which
    a complete, self-contained program that performs a specific function
    directly for the user is expressed or is running.}.
These faults, possibly having their origin at design time, or during operation, or
while interacting with the environment, are not different in the extent of their
consequences from those faults originating, e.g., in the hardware or the operating system.
A well known example of this is the case of the Ariane 5 flight 501~\cite{IBR96}, in which 
the consequences of a fault in the application ultimately brought to a system crash.
In general we observe how
the higher the level of abstraction, 
the higher the complexity of the algorithms into play 
and the consequent error proneness of the involved (real or abstract) machines.
As a conclusion, full tolerance of faults and complete fulfilment
of the dependability design goals of a complex software application
must include means to avoid, remove, or tolerate faults
working at all levels, including the application layer. 
This paper focuses
on run-time detection and recovery of faults through mechanisms residing in
or cooperating with the application layer.
\subsection{Software Fault-Tolerance in the Application Layer}\label{S:alft}
The need of software fault-tolerance provisions, located
in the application layer, is supported by studies that showed that the
majority of failures experienced by modern computer systems are due to
software faults, including those located in the application
layer~\cite{Lyu98a,Lyu98b,ALRL04,ALR04}; for instance, 
NRC reported that 81\% of the total number of outages of US switching systems
in 1992 were due to software faults~\cite{NRC93}. Moreover,
modern application software systems are
increasingly networked and distributed. Such systems, e.g., client-server
applications, are often characterised by a loosely coupled architecture whose
global structure is in general more prone to failures\footnote{As Leslie Lamport
	efficaciously synthesised in his quotation, ``a distributed system 
	is one in which I cannot get something done because a machine I've never heard of 
	is down''.}.
Due to the complex and temporal nature of interleaving of messages and
computations in distributed software systems, no amount of verification,
validation and testing can eliminate all faults in an application and give
complete confidence in the availability and data consistency of applications
of this kind~\cite{HuKi95}. Under these assumptions, the only alternative
(and effective) means for increasing software reliability is that of
incorporating in the application software provisions of software fault-tolerance~\cite{Ran75}.

Another argument that justifies the addition of software fault-tolerance means
in the application layer is given by the  widespread adoption of reusable software
components. Approaches such as object-orientation, component-based middleware and
service-orientation have provided the designer with effective tools to compose
systems out of e.g., COTS object libraries, third-party components, and web services.
For instance, many object-oriented applications are indeed built from
reusable components
the sources of which are unknown to the application developers.
The above mentioned approaches
fostered the capability of dealing with higher levels of complexity in software
and at the same time eased and therefore encouraged software reuse. This is having
a big, positive impact on development costs, but turns the application 
into a sort of collection of reused, pre-existing ``blocks'' made by 
third parties. The reliability of these components and therefore their impact on the
overall reliability of the user application is often unknown, up to the
point that
Green defines as ``art'' creating reliable applications
using off-the-shelf software components~\cite{Gre97}. The case
of the Ariane 501 flight is a well-known example that shows how improper
reuse of software may have severe consequences%
\footnote{The Ariane 5\label{foot}
	programme reused the long-tested software used
	in Ariane 4. Such software
	had been thoroughly tested and was
	compliant to Ariane 4 specifications. Unfortunately, specifications
	for Ariane 5 were different.
	A dormant design fault
	had never been unravelled simply because the operating conditions of Ariane 4
	were different from those of Ariane 5.
	This failure entailed
	a loss of about 370
	million Euros~\cite{LeLa96}.}~\cite{IBR96}.

But probably the most convincing argument for not excluding the
application layer from a fault-tolerance strategy is the
so-called ``end-to-end argument'', a system design principle
introduced by Saltzer, Reed and Clark~\shortcite{SaRC84}.
This principle states that, rather often, functions such as
reliable file transfer, can be completely and correctly 
implemented only with the knowledge and help of
the application standing at the endpoints of the underlying system
(for instance, the communication network).

This does not mean that
everything should be done at the application level---fault-tolerance strategies
in the underlying hardware and operating system can have a strong
impact on a system's performance. However, an extraordinarily
reliable communication system, that guarantees that no packet is lost,
duplicated, or corrupted, nor delivered to the wrong addressee, does not
reduce the burden of the application program to ensure reliability:
for instance, for reliable file transfer, the application programs
that perform the transfer must still supply a file-transfer-specific,
end-to-end reliability guarantee.

Hence one can conclude that:
\begin{quote}
Pure hardware-based or operating 
system-based solutions to fault-to\-le\-ran\-ce, though often characterised by 
a higher degree of transparency, are not fully capable of providing 
complete end-to-end tolerance to faults in the user application. 
Furthermore, relying solely on the hardware and the operating system
develops only partially satisfying solutions;
requires a large amount of extra resources and costs;
and is often characterised by poor service portability~\cite{SaRC84,SiSw92}.
\end{quote}

\subsection{Strategies, Problems, and Key Properties}\label{Sect:StrAttr}
The above conclusions justify the strong need for ALFT;
as a consequence of this need, several approaches
to ALFT have been devised during the last three decades
(see Chapter~\ref{Chap:DesignTools} for a brief survey).
Such a long research period hints at the complexity
of the design problems underlying ALFT engineering,
which include:
\begin{enumerate}
\item How to incorporate fault-tolerance in the application layer
 of a computer program. 
\item Which fault-tolerance provisions to support.
\item How to manage the fault-tolerant code.
\end{enumerate}

Problem 1 is also known as the problem of the
\textbf{system structure to software fault-tolerance}, first
proposed by B. Randell in~\shortcite{Ran75}. It states the need of appropriate
structuring techniques such that the incorporation
of a set of fault-tolerance provisions in the application software
might be performed in a simple, coherent, and well structured way.
Indeed, poor solutions to this problem result in a huge degree of
\textbf{code intrusion}: 
in such cases, the application code that addresses the functional requirements
and the application code that addresses the fault-tolerance requirements
are mixed up into one large and complex application software.
\begin{itemize}
\item This greatly complicates the task of the developer and requires expertise
  in both the application domain and in fault-tolerance. Negative
  repercussions on the development times and costs are to be expected.
\item The maintenance of the resulting code, both for
  the functional part and for the fault-tolerance provisions,
  is more complex, costly, and error prone. 
\item Furthermore, the overall complexity of the software product is
  increased---which is detrimental to its resilience to faults.
\end{itemize}
One can conclude that, with respect to the first problem, an ideal
system structure should guarantee an adequate
\textbf{Separation between the functional and
the fault-tolerance Concerns} (\SC\label{Def:SC}).

\vspace*{5pt}

Moreover, the design choice of
which fault-tolerance provisions to support
can be conditioned by the adequacy of the syntactical
structure at ``hosting'' the various provisions.
The well-known quotation by B. L. Whorf efficaciously
captures this concept:
\begin{quote}\label{Pageref:Whorf}
``Language shapes the way we think,
  and determines what we can think about.''
\end{quote}
Indeed, as explained in Sect.~\ref{Sect:Paradox},
a non-optimal answer to Problem 2 may
\begin{itemize}
\item require a high degree of redundancy, and
\item rapidly consume large amounts of the available
      redundancy,
\end{itemize}
which at the same time would increase the costs and
reduce reliability.
One can conclude that, devising a syntactical structure offering
straightforward support to
a large set of fault-tolerance provisions,
can be an important aspect of an ideal
system structure for ALFT.
In the following this property will be called
\textbf{Syntactical Adequacy} (\SA\label{Def:SA}).

\vspace*{5pt}

Finally, one can observe that another
important aspect of an ALFT architecture is
the way the fault-tolerant code is managed,
at compile time as well as at run time. Evidence for
this statement can be found by observing how
\label{Def:EMI}
a number of important choices pertaining to the
      adopted fault-tolerance provisions, such as
      the parameters of a temporal redundancy strategy,
      are a consequence of an analysis of the environment
      in which the application is to be deployed and is
      to run\footnote{For instance, if an application is to be moved from a
         domestic environment to another one characterised by an
         higher electro-magnetic interference (EMI),
         it is reasonable to assume that, e.g., the number of
         replicas of some protected resource should be
         increased accordingly.}.
      In other words, depending on the target environments,
      the set of (external) impairments that
      might affect the application can vary considerably.
      Now, while it may be
      in principle straightforward to port an existing code
      to another computer system, 
      \textbf{porting the service} supplied by that code may require
      a proper adjustment of the above mentioned choices, namely
      the parameters of the adopted provisions~\cite{DB07c}.
      Effective support towards the management of the parametrisation
      of the fault-tolerant code and, in general, of its
      maintenance, could guarantee \textbf{fault-tolerance software reuse}.
Therefore, off-line and on-line (dynamic) management
of fault-tolerance provisions and their parameters may be
an important requirement for any satisfactory solution of Problem 3.
As further motivated in Sect.~\ref{Sect:Paradox},
ideally, the fault-tolerant code should \emph{adapt\/} itself
to the current environment. Furthermore, any satisfactory
management approach should not overly increase the complexity
of the application---which would be detrimental to dependability.
Let us call this property \textbf{Adaptability} (\Aty\label{Def:A}).

Let us refer collectively to properties \SC, \SA{} and \Aty{} as to the
\emph{structural attributes\/} of ALFT.


\vspace*{5pt}

The various approaches to ALFT surveyed in Section~\ref{Chap:DesignTools}
provide different system structures to solve the above mentioned problems.
The three structural attributes are used in that section in order to provide
a qualitative assessment with respect to various application requirements.
The structural attributes constitute, in a sense, a \emph{base\/}
with which to perform this assessment.
One of the major conclusions of that survey is that none of the
surveyed approaches is capable to provide the best combination of values
of the three structural attributes in every application domain.
For specific domains,
such as object-oriented distributed applications, satisfactory solutions
have been devised at least for \SC{} and \SA, while only partial solutions
exist, for instance, when dealing with the class of distributed or
parallel applications not based on the object model.

The above matter of facts has been efficaciously captured by Lyu,
who calls this situation
``the software bottleneck'' of system development~\cite{Lyu98b}:
in other words, there is evidence of an urgent need for 
systematic approaches to assure software reliability within
a system~\cite{Lyu98b}
while effectively addressing the above problems.
In the cited paper, Lyu remarks how ``developing the 
required techniques for software reliability engineering is a major challenge
to computer engineers, software engineers and engineers of related disciplines.''



\subsubsection{Fault-Tolerance, Redundancy, and Complexity}\label{Sect:Paradox}
A well-known result by Shannon~\shortcite{Shannon} tells us that,
from any unreliable channel, it is possible to set up
a more reliable channel by increasing the degree
of information redundancy. This means that
it is possible to trade off reliability and redundancy
of a channel.
The authors observe that
the same can be said for a fault-tolerant system,
because fault-tolerance is in general the result
of some strategy effectively exploiting some form
of redundancy---time, information, and/or
hardware redundancy~\cite{John89a}.
This redundancy has a cost penalty attached, though.
Addressing a weak failure semantics, able to span
many failure behaviours, effectively translates
into higher reliability---nevertheless,
\begin{enumerate}
\item it \textbf{requires} large amounts of extra resources,
      and therefore implies a high cost penalty, and
\item it \textbf{consumes} large amounts of extra resources,
      which translates into the rapid exhaustion of the
      extra resources.
\end{enumerate}
For instance, 
Lamport \emph{et al}.~\shortcite{LaSP82} set the minimum level of redundancy
required for tolerating Byzantine failures to a value that is
greater than the one required for tolerating, e.g., value failures.
Using the simplest of the algorithms described in the cited paper,
a 4-modular-redundant (4-MR) system can only withstand
any single Byzantine failure, while the same system
may exploit its redundancy to withstand up to
three crash faults---though no other kind of
fault~\cite{Pow97a}. In other words:
\begin{quote}
After the occurrence of a crash fault,
a 4-MR system with strict Byzantine failure semantics has exhausted its
redundancy and is no more dependable than a non-redundant system supplying
the same service, while the crash failure semantics system is able to survive
to the occurrence of that and two other crash faults. On the other hand, the
latter system, subject to just one Byzantine fault, would fail regardless
its redundancy.
\end{quote}

We can conclude that for any given level of redundancy
trading complexity of failure mode against number and type
of faults tolerated may be an important capability
for an effective fault-tolerant structure. Dynamic adaptability
to different environmental conditions\footnote{%
           The following quote by J. Horning~\shortcite{Hor98} captures
           very well how relevant may be the role of the environment
           with respect to achieving the required quality of service:
           ``What is the most often overlooked risk in software engineering?
           That the environment will do something the designer never
           anticipated''.}
may provide a satisfactory answer
to this need, especially when the additional complexity does
not burden (and jeopardise) the application.
Ideally, such complexity should be part of a custom architecture
and not of the application.
On the contrary, the embedding in the application of complex failure semantics,
covering many failure modes, implicitly promotes complexity, as it may
require the implementation of many recovery mechanisms. This complexity is
detrimental to the dependability of the system, as it is in itself a
significant source of design faults. Furthermore, the isolation
of that complexity outside the user application may allow cost-effective
verification, validation and testing. These processes may be unfeasible
at the application level.

The authors conjecture that\label{conj2}
a satisfactory solution to the design problem of
the management of the fault-tolerant code
(presented in Sect.~\ref{Sect:StrAttr})
may translate into an optimal management of the
failure semantics (with respect to the
involved penalties). In other words, we conjecture
that 
linguistic structures characterised by high
\emph{adaptability\/} (\Aty) may be better suited to cope
with the just mentioned problems.

\section{Current Approaches to Application-Level 
	 Fault-Tolerance}\label{Chap:DesignTools}
One of the conclusions drawn in
Sect.~\ref{Chap:Intro} is that the system to be made fault-tolerant
must include provisions for fault-tolerance 
also in the application layer of a computer program.
In that context, the problem of which system structure to use
for ALFT has been proposed. 
This section provides a critical survey of the state-of-the-art 
on embedding fault-tolerance means in the application layer.

According to the literature, at least six classes of methods,
or approaches, can be used for embedding fault-tolerance provisions
in the application layer of a computer program. This section
describes these approaches and points out positive and negative
aspects of each of them
with respect to the structural attributes defined in
Sect.~\ref{Sect:StrAttr} and to various application domains.
A non-exhaustive list of the systems and projects implementing 
these approaches is also given.
Conclusions are drawn in Sect.~\ref{Chap:end}, where the need 
for more effective approaches is recognised.

Two of the above mentioned approaches derive from well-established
research in software fault-tolerance---Lyu~\shortcite{Lyu98b,Lyu96,Lyu95} refers
to them as single-version and multiple-version software fault-tolerance.
They are dealt with in Sect.~\ref{Sect:NVersion}.
A third approach, described in Sect.~\ref{Sect:MOP}, is based on 
metaobject protocols. It is derived from the domain 
of object-oriented design and can also be used for embedding services 
other than those related to fault-tolerance. A fourth approach
translates into developing new custom high-level distributed
programming languages or enhancing pre-existent languages of that kind.
It is described in Sect.~\ref{Sect:Lang}. 
Aspect-oriented programming as a fault-tolerance structuring technique
is discussed in Sect.~\ref{Sect:AOP}.
Finally, Sect.~\ref{Sect:RMP}
describes an approach, based on a special recovery task monitoring
the execution of the user task.

\subsection{Single- and Multiple-Version Software Fault-Tolerance}\label{Sect:NVersion}
A key requirement for the development of fault-tolerant systems is the availability
of \textbf{replicated resources}, in hardware or software. A 
fundamental method employed to attain fault-tolerance is \textbf{multiple computation}, 
i.e., $N$-fold ($N\ge2$) replications in three domains:
\begin{description}
\item[Time] That is, repetition of computations.
\item[Space] I.e., the adoption of multiple hardware channels (also called ``lanes'').
\item[Information] That is, the adoption of multiple versions of software.
\end{description}
Following Avi\u{z}ienis~\shortcite{Avi85}, it is possible to characterise at least
some of the approaches towards fault-tolerance by means of a notation
resembling the one used to classify queueing systems models~\cite{Kle75}:
\[ \ths nmp, \]
\label{func:ths}%
the meaning of which is ``$n$ executions, on $m$ hardware channels, of $p$ programs''.
The non-fault-tolerant system, or \ths111, is called \emph{simplex\/} in the
cited paper.

\subsubsection{Single-Version Software Fault-Tolerance}\label{SubSect:SVSFT}
Single-version software fault-to\-le\-ran\-ce (SV) is basically the embedding
into the user application of a simplex system of error detection
or recovery features, e.g., atomic actions~\cite{JaCa85},
checkpoint-and-rollback~\cite{EAW02}, or exception handling~\cite{Cri95}.
The adoption of SV in the application layer requires
the designer to concentrate in one physical location,
namely the source code of the application, both the specification
of what to do in order to perform some user computation and
the strategy such that faults are tolerated when they occur.
As a result, the size of the problem addressed is increased. \emph{A fortiori},
this translates into an increase of size of the user application, which
induces loss of transparency, maintainability, and
portability while increasing development times and costs. 

A partial solution to this loss in portability and these higher costs
is given by the development of libraries and frameworks created
under strict software engineering processes. In the following, two examples
of this approach are presented---the EFTOS library and the SwiFT system.

\paragraph{The EFTOS library}\label{Sect:EFTOS}
EFTOS~\cite{DeDL98a} (that is ``Embedded, Fault-Tolerant 
Supercomputing'') is the name of ESPRIT-IV project 21012. Aims of this project
were to investigate approaches to add fault-tolerance to embedded high-performance
applications in a flexible and cost-effective way.
The EFTOS library has been first implemented on a Parsytec 
CC system~\cite{Anon96a}, 
a distributed-memory MIMD supercomputer consisting of processing 
nodes based on PowerPC 604 microprocessors at 133MHz, 
dedicated high-speed links, I/O modules,
and routers.

\begin{figure}
\centerline{\includegraphics[width=14cm]{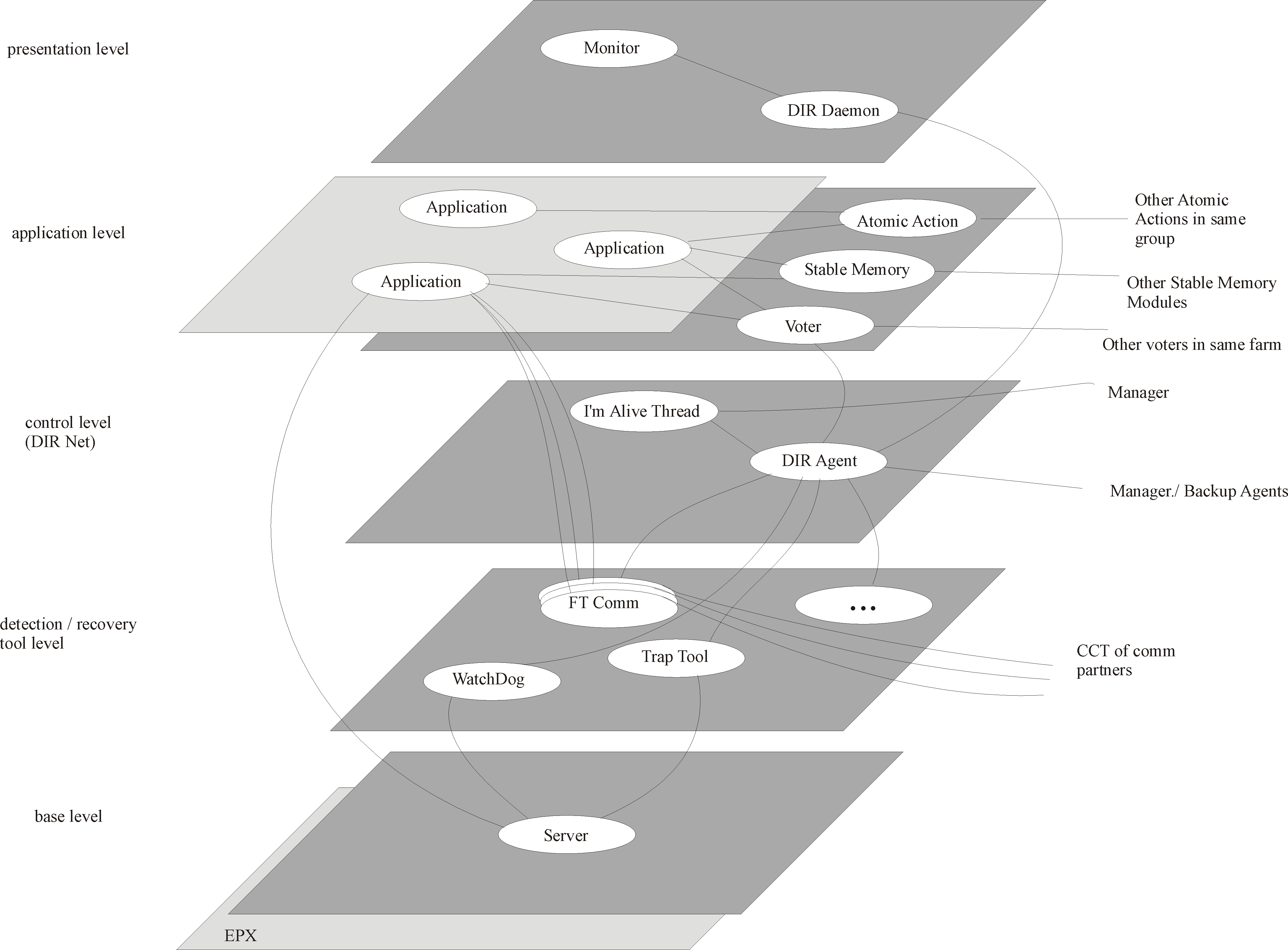}}
\caption{The structure of the EFTOS library. Light gray has been used
for the operating system and the user application, while dark gray layers
pertain EFTOS.}
\label{Fig:library}
\end{figure}

Through the adoption of the EFTOS library, the target embedded parallel 
application is plugged into a hierarchical, layered system whose
structure and basic components (depicted in Fig.~\ref{Fig:library}) are:
\begin{itemize}
\item At the base level, a distributed net of ``servers'' whose
main task is mimicking possibly missing (with respect to the POSIX
standards) operating system functionalities, such as remote thread creation;

\item One level upward (detection tool layer), a set of parameterisable functions managing error
detection, referred to as ``Dtools''.
These basic components are plugged into the embedded
application to make it more dependable. EFTOS supplies a number of these
Dtools, e.g., a watchdog timer thread and a trap-handling
mechanism, plus an API for incorporating user-defined EFTOS-com\-pli\-ant tools; 

\item At the third level (control layer), a distributed application
called ``DIR net'' (its name stands for ``detection, isolation,
and recovery network'') is used to coherently combine the Dtools,
to ensure consistent fault-tolerance strategies throughout
the system, and to play the role of a backbone handling information
to and from the fault-tolerance elements~\cite{DeDL00d,DeFl98a}.
The DIR net can be regarded as a fault-tolerant network of
crash-failure detectors connected to other peripheral error detectors.
Each node of the DIR net is ``guarded'' by an \ImAlive{} thread
that requires the local component to send periodically ``heartbeats''
(signs of life). A special component, called RINT, manages error
recovery by interpreting a custom language called RL~\cite{DeDe02a,DeFl97c}.

\item At the fourth level (application layer), the Dtools and the components of
the DIR net are combined into dependable mechanisms i.e.,
methods to guarantee fault-tolerant communication~\cite{EVMV98}, 
tools implementing a virtual Stable Memory~\cite{PDP2001},
a distributed voting mechanism called 
``voting farm''~\cite{DeFl97b,DeDL98a,DeDL98e}, and so forth;

\item The highest level (presentation layer) is given by
a hypermedia distributed application
which monitors the structure 
and the state of the user application~\cite{DeDe98}. This application is
based on a special CGI script~\cite{Kim96}, called DIR Daemon, which 
continuously takes its inputs from the DIR net, 
translates them into HTML,
and remotely controls a Netscape browser~\cite{Zaw} so that it renders
these HTML data.
\end{itemize}

\paragraph{The SwiFT System}\label{Pag:SwiFT}
SwiFT~\cite{HuKi+96} stands for Software Implemented Fault-Tolerance.
It includes a set of reusable software components
(\texttt{watchd}, a general-purpose UNIX daemon watchdog timer; \texttt{libft}, a
library of fault-tolerance methods, including single-version
implementation of recovery blocks and $N$-version programming (see
Sect.~\ref{SubSect:MVSFT}); \texttt{libckp}, i.e., a user-transparent 
checkpoint-and-rollback
library; a file replication mechanism called \texttt{REPL};
and \texttt{addre\-juv}, a special ``reactive'' feature of \texttt{watchd}~\cite{HuKi95+}
that allows for software rejuvenation\footnote{Software
   rejuvenation~\cite{HuKi95+,BST03} offers tools for periodical and graceful
   termination of an application with immediate restart, so that possible
   erroneous internal states, due to transient faults, be wiped out before
   they turn into a failure.}.
SwiFT has been successfully used and proved to be
efficient and economical means to increase the level of fault-tolerance
in a software system where residual faults are present and their
toleration is less costly than their full elimination~\cite{Lyu98b}.
A relatively small overhead is introduced in most cases~\cite{HuKi95}.

%

\paragraph{Conclusions}
\begin{figure}
\centerline{\includegraphics[width=5cm]{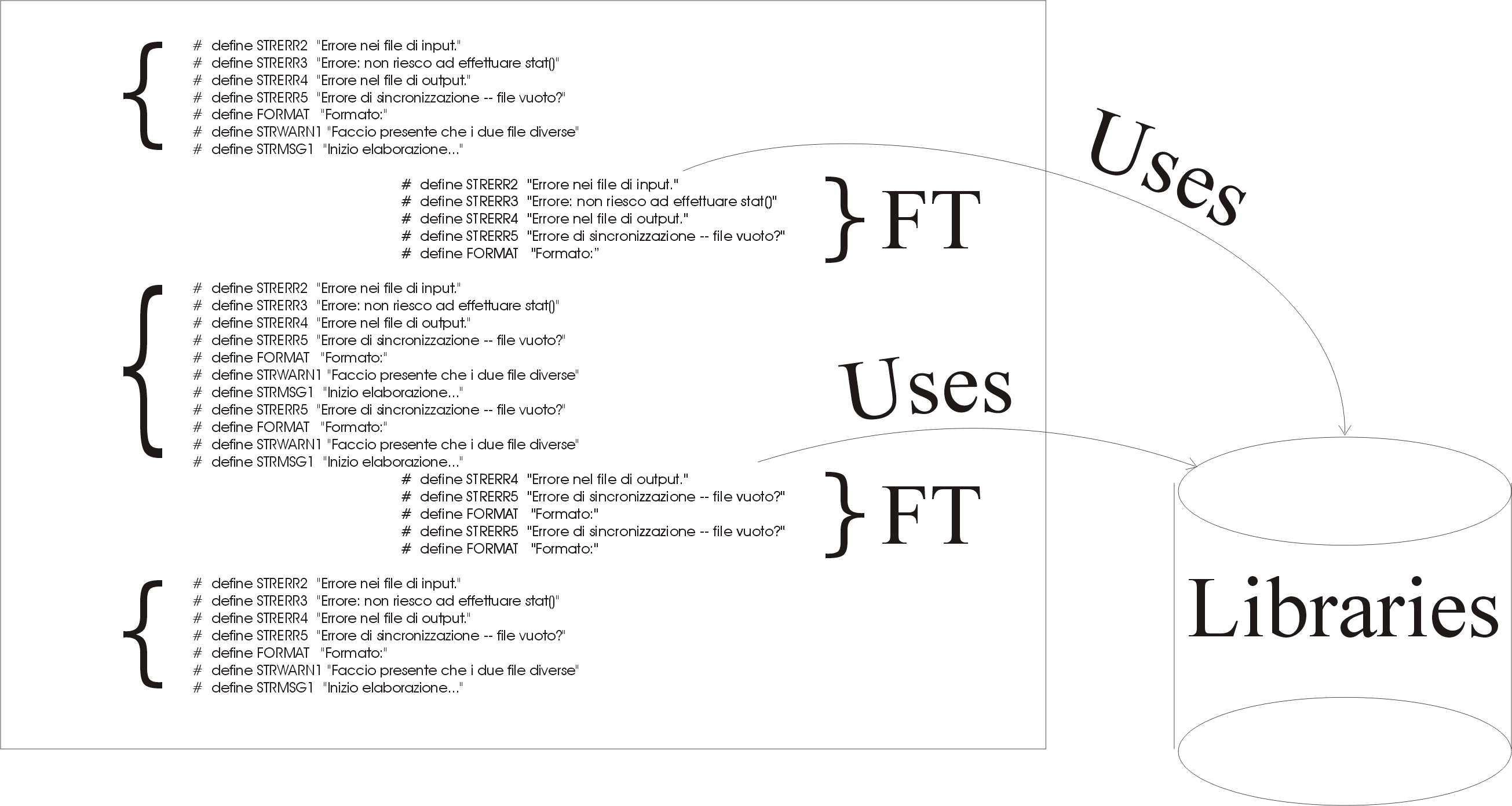}}
\caption{A fault-tolerant program according to a SV system.}\label{f:phi-SV}
\end{figure}
Figure~\ref{f:phi-SV} synthesizes the main characteristics of
the SV approach: the functional and the fault-tolerant code are
intertwined and the developer has to deal with the two concerns at 
the same time, even with the help of libraries of fault-tolerance
provisions.
In other words, SV requires
the application developer to be an expert in fault-tolerance as well,
because he (she) has to integrate in the application a number of
fault-tolerance provisions among those available
in a set of ready-made basic tools. His (hers) is the responsibility
for doing it in a coherent, effective, and efficient way.
As it has been observed in Sect.~\ref{Sect:StrAttr},
the resulting code is a mixture of functional code and of custom 
error-management code that does not always offer an acceptable degree
of portability and maintainability.
The functional and non-functional design concerns are not kept apart
with SV, hence one can conclude that (qualitatively)
SV exhibits poor separation of concerns (\SC).
This in general has a bad impact on design and maintenance \textbf{costs}.

As to syntactical adequacy (\SA),
we observe that following SV the fault-tolerance provisions are offered
to the user though an interface based on a general-purpose
language such as C or \CPP. As a consequence, very limited \SA{}
can be achieved by SV as a system structure for ALFT.

Furthermore, no support is provided
for off-line and on-line configuration of the fault-tole\-ran\-ce provisions.
Consequently we regard the adaptability (\Aty) of this approach as insufficient.

On the other hand, tools in SV libraries and systems
give the user the ability to deal with fault-tolerance ``atoms''
without having to worry about their actual implementation and with a good
ratio of costs over improvements of the dependability
attributes, sometimes introducing a relatively small overhead.
Using these toolsets the designer can re-use existing, long tested,
sophisticated pieces of software without having each time 
to ``re-invent the wheel''. 

Finally, it is important to remark that, in principle, SV poses
no restrictions on the class of applications that may be
tackled with it.

\subsubsection{Multiple-Version Software Fault-Tolerance}\label{SubSect:MVSFT}
This section describes multiple-version software fault-tolerance
(MV), an approach which requires
$N$ ($N\ge2$) independently designed versions of software. MV
systems are therefore \ths xyN systems. 
In MV, a same service or functionality is supplied
by $N$ pieces of code that
have been designed and developed by different, 
independent software teams\footnote{This requirement is well explained
by Randell~\shortcite{Ran75}:
	``All fault-tolerance must be based on the provision of useful
	redundancy, both for error detection and error recovery. In
	software the redundancy required is not simple replication of
	programs but \emph{redundancy of design}.''
	Footnote~\ref{foot} briefly reports on the 
	consequences of a well known case of redundant design.}.
The aim of this approach is to reduce the effects of design faults
due to human mistakes committed at design time.
The most used configurations are \ths N1N, i.e., $N$ sequentially applicable
alternate programs using the same hardware channel, and \ths 1NN, based on
the parallel execution of the alternate programs on $N$, possibly diverse,
hardware channels.

Two major approaches exist:
the first one is known as recovery block~\cite{Ran75,RaXu95},
and is dealt with in Sect.~\ref{Sect:RecoveryBlock}. The second approach
is the so-called $N$-version programming~\cite{Avi85,Avi95}.
It is described in Sect.~\ref{Sect:NVersProg}.

\paragraph{The Recovery Block Technique}\label{Sect:RecoveryBlock}
Recovery Blocks are usually implemented as \ths N1N{} systems.
The technique addresses residual software design faults. It aims
at providing fault-tolerant functional components which may be nested
within a sequential program. Other versions of the approach,
implemented as \ths 1NN{} systems, are suited for parallel
or distributed programs~\cite{Sco85,RaXu95}.

\begin{figure}
\centerline{\includegraphics[width=14cm]{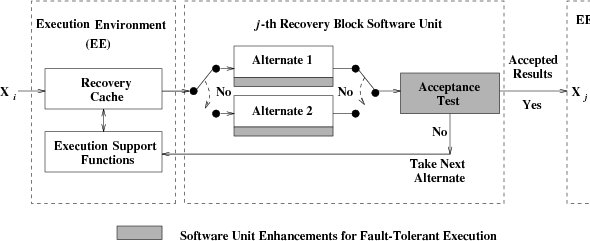}}
\caption{The recovery block model with two alternates. The execution
environment 
is charged with the management of the recovery cache and the execution support
functions (used to restore the state of the application when the acceptance test
is not passed), while the user is responsible for supplying both alternates and
the acceptance test.}
\label{Fig:RecoveryBlock}
\end{figure}
The recovery blocks technique is similar to the hardware fault-tolerance 
approach known as ``stand-by sparing'', which is described, e.g., in~\cite{John89a}.
The approach is
summarised in Fig.~\ref{Fig:RecoveryBlock}: on entry to a recovery block,
the current state of the system is checkpointed. A primary alternate is
executed.
When it ends, an acceptance test checks whether the primary alternate 
successfully accomplished its objectives. If not, a backward recovery step
reverts the system state back to its original value and a secondary alternate
takes over the task of the primary alternate. When the secondary alternate
ends, the acceptance test is executed again. The strategy goes on until
either an alternate fulfils its tasks or all alternates are executed
without success. In such a case, an error routine is executed.
Recovery blocks can be nested---in this case the error routine invokes
recovery in the enclosing block~\cite{RaXu95}. An exception triggered within
an alternate is managed as a failed acceptance test.
A possible syntax for recovery blocks is as follows:

\begin{verbatim}
     ensure       acceptance test
     by           primary alternate
     else by      alternate 2
	          .
	          .
     else by      alternate N
     else error
\end{verbatim}

Note how this syntax does not explicitly show the recovery step that should
be carried out transparently by a run-time executive.

The effectiveness of recovery blocks rests to a great extent on the coverage of the error
detection mechanisms adopted, the most crucial component of which is the acceptance
test.
A failure of the acceptance test is a failure of the whole recovery blocks strategy. For this
reason, the acceptance test must be simple, must not introduce huge run-time
overheads, must not retain data locally, and so forth. It must be regarded
as the ultimate means for detecting errors, though not the
exclusive one.
Assertions and run-time checks, possibly supported by underlying layers,
need to buttress the strategy and reduce the probability of an acceptance
test failure.
Another possible failure condition for the recovery blocks approach is given by
an alternate failing to terminate. This may be detected by a time-out 
mechanism\label{time-out} that could be added to recovery blocks. This
addition obviously further increases the complexity.

The SwiFT library that has been described 
in Sect.~\ref{SubSect:SVSFT} (p.~\pageref{Pag:SwiFT}) implements
recovery blocks in the C language as follows:

\begin{verbatim}
     #include <ftmacros.h>
     ...
     ENSURE(acceptance-test) {
                primary alternate;
     } ELSEBY {
	        alternate 2;
     } ELSEBY {
	        alternate 3;
     }
     ...
     ENSURE;
\end{verbatim}

Unfortunately this approach does not cover any of the above mentioned requirements
for enhancing the error detection coverage of the acceptance test.
This would clearly require a run-time executive that is not part of this strategy.
Other solutions, based on enhancing the grammar of pre-existing programming
languages such as Pascal~\cite{Shr78} and Coral~\cite{And85}, have some impact
on portability.
In both cases, code intrusion is not avoided. 
This translates into difficulties when trying to modify or maintain
the application program without interfering ``much'' with the
recovery structure, and vice-versa, when trying to modify or
maintain the recovery structure without interfering ``much''
with the application program. Hence one can conclude that
recovery blocks are characterised by unsatisfactory values of the structural
attribute \SC.
Furthermore, a system structure for ALFT based exclusively on
recovery blocks does not satisfy attribute \SA\footnote{Randell
         himself states that, given the ever increasing complexity
         of modern computing, there is still 
         an urgent need for ``richer forms of structuring for error
         recovery and for design diversity''~\cite{RaXu95}.}.
Finally, regarding attribute
\Aty{}, one can observe that recovery blocks are
a rigid strategy that does not allow off-line configuration nor
(\emph{a fortiori}) code adaptability.

\vspace*{5pt}

On the other hand, 
recovery blocks have been successfully adopted throughout 30 years in many
different application fields. It has been successfully validated by a
number of statistical
experiments and through mathematical modelling~\cite{RaXu95}. Its adoption
as the sole fault-tolerance means, while developing complex applications,
resulted in some cases~\cite{And85} in a failure coverage of over 70\%,
with acceptable overheads in memory space and CPU time.

A negative aspect in any MV system is given by
development and maintenance \textbf{costs}, that
grow as a monotonic function of $x, y, z$ in any \ths xyz{} system.
\paragraph{$\mathbf N$-Version Programming}\label{Sect:NVersProg}
$N$-Version Programming (NVP) systems are built from
generic architectures based on redundancy and consensus.
Such systems usually belong to class
\ths 1NN, less often to class \ths N1N. NVP is defined by its
author~\cite{Avi85} as ``the independent 
generation of $N\ge2$ functionally equivalent programs from the same initial
specification.'' These $N$ programs, called versions, are developed
for being executed in parallel. This system constitutes a fault-tolerant
software unit that depends on a generic decision algorithm to determine a
consensus or majority result from the individual outputs of two or more
versions of the unit.

\begin{figure}
\centerline{\includegraphics[width=13cm]{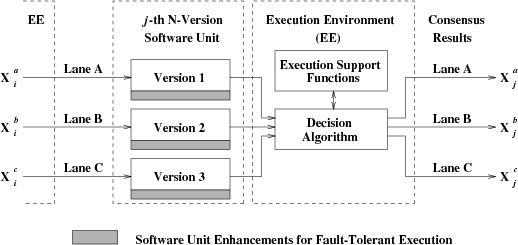}}
\caption{The $N$-Version Software model when $N=3$. The execution environment 
is charged with the management of the decision algorithm and with the execution support
functions. The user is responsible for supplying the $N$ versions. Note how
the Decision Algorithm box takes care also of multiplexing its output onto
the three hardware channels---also called ``lanes''.}
\label{Fig:NVersProg}
\end{figure}

Such a strategy
(depicted in Fig.~\ref{Fig:NVersProg}) has been developed under
the fundamental conjecture that independent designs translate into
random component failures---i.e., statistical independence.
Such a result would guarantee that correlated failures do not translate
into immediate exhaustion of the available redundancy, as it would happen, e.g.,
using $N$ copies of the same version. Replicating software would also mean
replicating any dormant software fault in the source version---see, e.g.,
the accidents with the Therac-25 linear accelerator~\cite{Lev95}
or the Ariane 5 flight 501~\cite{IBR96}.
According to Avi\u{z}ienis, independent generation of the versions significantly
reduces the probability of correlated failures. Unfortunately a number of 
experiments~\cite{ECK91} and theoretical studies~\cite{EL85}
have shown that this assumption is not always correct.

The main differences between recovery blocks and NVP are:
\begin{itemize}
  \item Recovery blocks (in its original form) is a sequential strategy whereas NVP allows
    concurrent execution;
  \item Recovery blocks require the user to provide a fault-free, application-specific,
    effective acceptance test, while NVP adopts a generic consensus
    or majority voting algorithm that can be provided
    by the execution environment (EE);
  \item Recovery blocks allow different correct outputs from the alternates, while the
    general-purpose character of the consensus algorithm of NVP calls for a single
    correct output\footnote{This weakness
      of NVP can be narrowed, if not solved, adopting the approach used in the
      so-called ``voting farm''~\cite{DeDL98e,DeDL98a,DeFl97b}, a generic
      voting tool designed by one of the authors of this paper in the
      framework of his participation to project ``EFTOS'' 
      (see Sect.~\ref{Sect:EFTOS}):
      such a tool works with opaque objects
      that are compared by means of a user-defined function.
      This function returns an integer value representing a ``distance''
      between any two objects to be voted. The user may choose between
      a set of predefined distance
      functions or may develop an application-specific distance function.
      Doing the latter, a distance may be endowed with the ability to
      assess that bitwise different objects are semantically
      equivalent. Of course, the user is still responsible for
      supplying a bug-free distance function---though is assisted
      in this simpler task by a number of template functions
      supplied with that tool.}.
\end{itemize}

The two models collapse when the acceptance test of recovery blocks is done as in NVP, i.e.,
when the acceptance test is a consensus on the basis of the outputs of the
different alternates.

\paragraph{Conclusions}
As with recovery blocks, also NVP has been successfully adopted for many years in various
application fields, including safety-critical airborne and spaceborne
applications. The generic NVP architecture, based on redundancy
and consensus, addresses parallel and distributed applications
written in any programming paradigm. A generic, parameterisable
architecture for real-time systems that supports 
the NVP strategy straightforwardly is GUARDS~\cite{Pow99}.

It is noteworthy to remark that the EE (also known as
$N$-Version Executive) is a complex component that needs
to manage a number of basic functions, for instance the execution
of the decision algorithm, the assurance of input consistency
for all versions, the inter-version communication, the version
synchronisation and the enforcement of timing constraints~\cite{Avi95}.
On the other hand, this complexity is not part of the application software---the $N$
versions---and it does not need to be aware of the fault-tolerance strategy.
An excellent degree of transparency can be reached, thus
guaranteeing a good value for attribute \SC.
Furthermore, as mentioned in Sect.~\ref{Sect:StrAttr},
costs and times required by a thorough verification, validation,
and testing of this architectural complexity may be acceptable,
while charging them to each application component is certainly not
a cost-effective option.

Regarding attribute \SA{}, the same considerations
provided when describing recovery blocks hold for NVP: also in this
case a single fault-tolerance strategy is followed.
For this reason we assess NVP as unsatisfactory regarding
attribute \SA.

Off-line adaptability to ``bad'' environments may be reached
by increasing the value of $N$---though this requires developing
new versions---a costly activity for both times and costs.
Furthermore, the architecture does not allow any dynamic
management of the fault-tolerance provisions. We conclude
that attribute \Aty{} is poorly addressed by NVP.

Portability is restricted by the portability
of the EE and of each of the $N$ versions. Maintainability actions may also
be problematic, as they need to be replicated and validated $N$ times---as
well as performed according to the NVP paradigm, so not to impact
negatively on statistical independence of failures. Clearly the same
considerations apply to recovery blocks as well. In other words, the adoption
of multiple-version software fault-tolerance provisions always implies
a penalty on maintainability and portability.

Limited NVP support has been developed for ``conventional'' programming
languages such as C. For instance, \texttt{libft}
(see Sect.~\ref{SubSect:SVSFT}, p.~\pageref{Pag:SwiFT}) implements
NVP as follows:

\begin{verbatim}
     #include <ftmacros.h>
     ...
     NVP
     VERSION{
                block 1;
                SENDVOTE(v_pointer, v_size);
     }
     VERSION{
                block 2;
                SENDVOTE(v_pointer, v_size);
     }
     ...
     ENDVERSION(timeout, v_size);
     if (!agreeon(v_pointer)) error_handler();
     ENDNVP;
\end{verbatim}

Note that this particular implementation extinguishes the potential transparency
that in general characterises NVP, as it requires some non-functional code
to be included. This translates into an unsatisfactory value for attribute \SC.
Note also that the execution of each block is in this case carried out
sequentially.

\vspace*{5pt}

It is important to remark how the adoption of NVP as a system
structure for ALFT requires a substantial increase in development and
maintenance \textbf{costs}: both \ths 1NN{} and  \ths N1N{}
systems have a cost function growing quadratically with $N$.
The author of the NVP strategy remarks how such costs
are paid back by the gain in trustworthiness.
This is certainly true when dealing with systems possibly subjected to
catastrophic failures---let us recall once more the case of the Ariane 5
flight 501~\cite{IBR96}. Nevertheless, the risks related to the chances
of rapid exhaustion of redundancy due to a burst of correlated failures
caused by a single or few design faults justify and call for the
adoption of other fault-tolerance provisions within and around the NVP unit
in order to deal with the case of a failed NVP unit.

\begin{figure}
\centerline{\includegraphics[width=9cm]{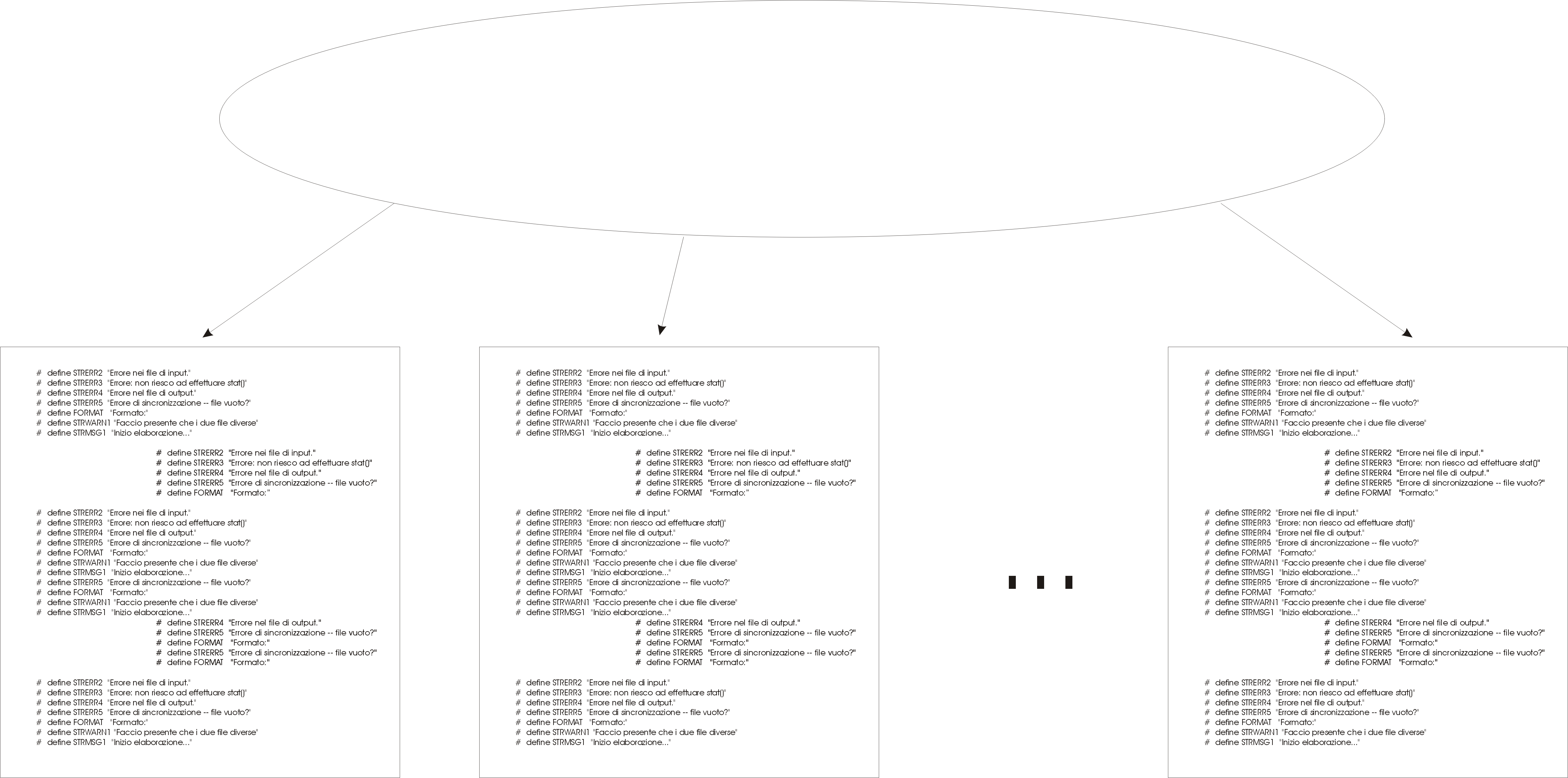}}
\caption{A fault-tolerant program according to a MV system.}\label{f:phi-MV}
\end{figure}
Figure~\ref{f:phi-MV} synthesizes the main characteristics of
the MV approach: several replicas of (portions of) the functional code
are produced and managed by a control component. In recovery blocks
this component is often coded side by side with the functional code
while in NVP this is usually a custom hardware box.
\subsubsection{A hybrid case: Data Diversity}
A special, hybrid case is given by data diversity~\cite{AmKn88}. A data diversity
system is a \ths 1N1 (less often a \ths N11). It can be concisely described as an NVP system
in which $N$ equal replicas are used as versions, but each replica receives a different minor
perturbation of the input data. Under the hypothesis that the function computed by the replicas
is non chaotic, that is, it does not produce very different output values when fed with slightly
different inputs, data diversity may be a cost-effective way to fault-tolerance. Clearly
in this case the voting mechanism does not run a simple majority voting but some vote fusion
algorithm~\cite{LoCE89}.
A typical application of data diversity is that of real time control programs, where sensor re-sampling
or a minor perturbation in the sampled sensor value may be able to prevent a failure.
Being substantially an NVP system, data diversity reaches the same values for the
structural attributes. The greatest advantage of this technique is that of drastically
decreasing design and maintenance costs, because design diversity is avoided.

\subsection{Metaobject Protocols and Reflection}\label{Sect:MOP}
\begin{figure}
\centerline{\includegraphics[width=7cm]{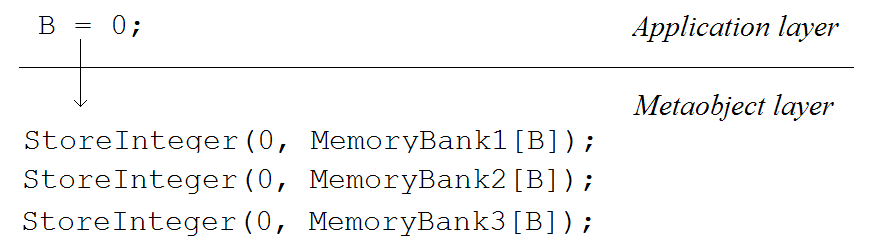}}

\vspace*{1cm}

\centerline{\includegraphics[width=7cm]{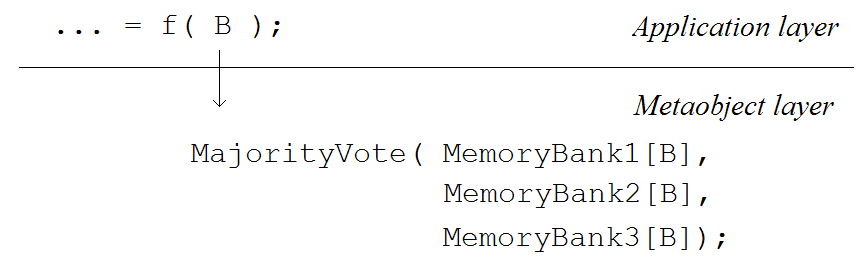}}
\caption{A MOP may be used to realize, e.g., triple-redundant memories
in a fully transparent way.}
\label{f:MOP}
\end{figure}

Some of the negative aspects pointed out while describing
single and multiple version software approaches
can be in some cases weakened, if not solved, by means of
a generic structuring technique which allows one to reach in some
cases an adequate degree of flexibility, 
transparency, and separation of design concerns: the adoption of 
metaobject protocols (MOPs)~\cite{KirB91}. The idea is to 
``open'' the implementation of the run-time executive of an object-oriented
language such as \CPP{} or Java so that
the developer can adopt and program different, custom semantics,
adjusting the language to the needs of the user and to the requirements
of the environment.
Using MOPs, the programmer can modify the behaviour of fundamental features such as
methods invocation, object creation and destruction, and 
member access.  The transparent management of spatial and temporal
redundancy~\cite{TaMB80} is a case where MOPs seem particularly adequate:
for instance, a MOP programmer may easily create 
``triple-redundant'' memory cells to protect his/her variables
against transient faults as depicted in Fig.~\ref{f:MOP}.

The key concept behind MOPs is that of computational reflection, or the
causal connection between a system and a meta-level description
representing structural and computational aspects of that system~\cite{Maes87}.
MOPs offer the metalevel programmer a representation of a system
as a set of metaobjects, i.e., objects that represent
and reflect properties of ``real'' objects, i.e., those objects that
constitute the functional part of the user application. Metaobjects
can for instance represent the structure of a class, or object
interaction, or the code of an operation. This mapping process
is called reification~\cite{Rob99}.

The causality relation of MOPs could also be extended to allow for
a dynamical reorganisation of the structure and the operation of a system, e.g.,
to perform reconfiguration and error recovery.
The basic object-oriented feature of inheritance
can be used to enhance the reusability of the FT mechanisms developed
with this approach.

\subsubsection{Project FRIENDS}
An architecture supporting this approach is the one developed in the
fra\-me\-work of
project FRIENDS~\cite{FaPe96a,FaPe98}.
The name FRIENDS stands for ``flexible and reusable implementation environment
for your next dependable system''. This project aims at implementing
a number of fault-tolerance provisions (e.g., replication, group-based
communication, synchronisation, voting\ldots~\cite{VAc97}) at meta-level.
In FRIENDS a distributed application is a set of objects interacting
via the proxy model, a proxy being a local intermediary between each object
and any other (possibly replicated) object.
FRIENDS uses the metaobject protocol provided by Open \CPP{}, a \CPP{}
preprocessor that provides control over instance creation and deletion,
state access, and invocation of methods.

Other ALFT architectures, exploiting the concept of metaobject
protocols within custom programming languages, are reported
in Sect.~\ref{Sect:Lang}.

\paragraph{Conclusions}
MOPs are indeed a promising system structure for embedding different
non-functional concerns in the application-level of a computer
program. MOPs work at \emph{language\/} level, providing a means
to modify the semantics of basic object-oriented language building
blocks such as object creation and deletion, calling and termination
of class methods, and so forth.
This appears to match perfectly
to a proper subset of the possible fault-tolerance provisions,
especially those such as transparent object redundancy that
can be straightforwardly managed with the metaobject approach.
When dealing with these fault-tolerance provisions, MOPs
provide a perfect separation of the design concerns, i.e.,
optimal \SC.
Some other techniques, specifically those who might be described as ``the
most coarse-grained ones'', such as distributed recovery blocks~\cite{Kim89},
appear to be less suited for being efficiently implemented via MOPs.
These techniques work at a distributed, macroscopic level.

\begin{figure}
\centerline{\includegraphics[width=10cm]{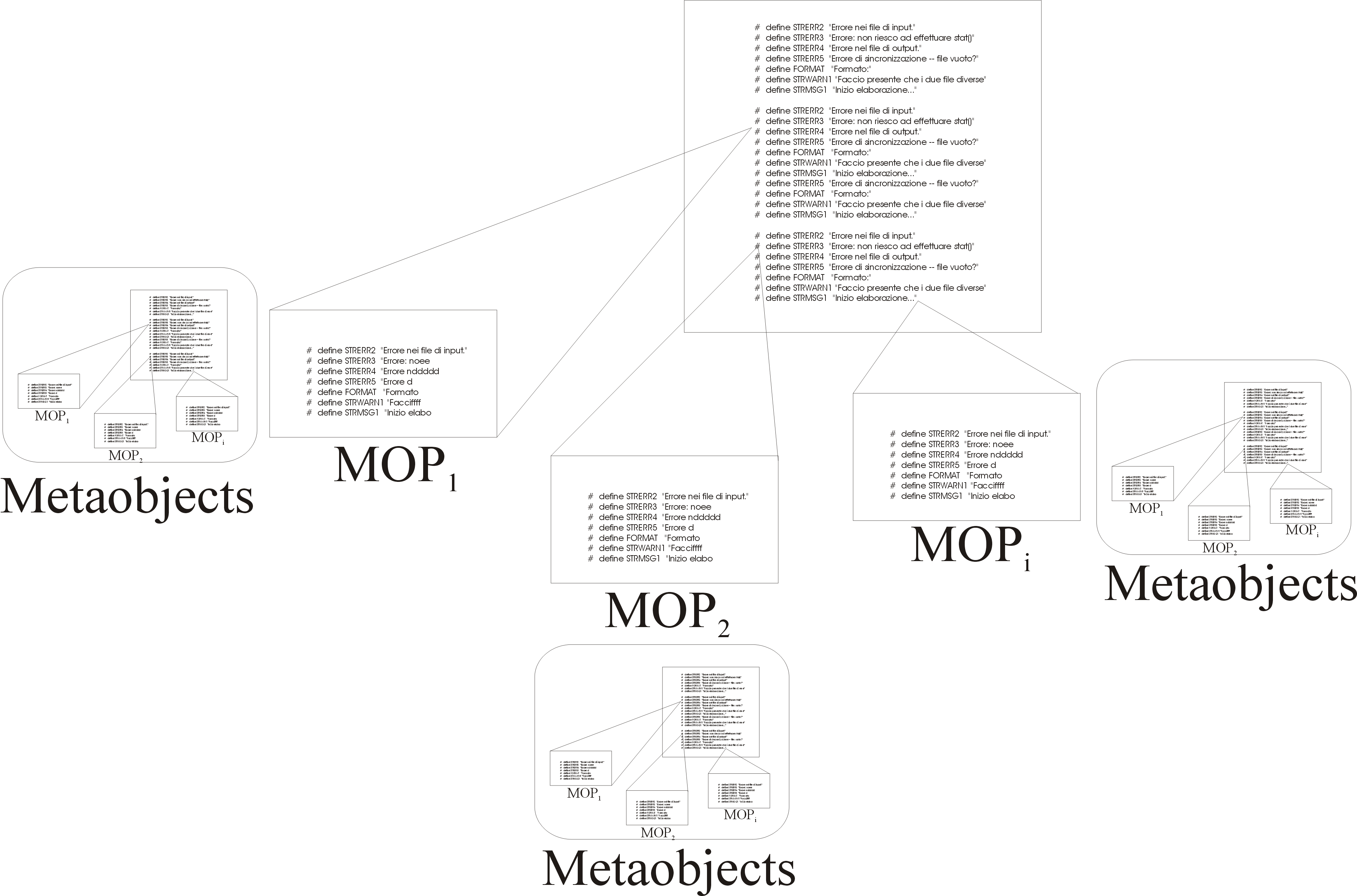}}
\caption{A fault-tolerant program according to a MOP system.}\label{f:phi-mop}
\end{figure}
Figure~\ref{f:phi-mop} synthesizes the main characteristics of
the MOP approach: The fault-tolerance programmer defines a number of
metaobject protocols and associates them with method invocations or
other grammar cases. Each time the functional program enters a certain
grammar case, the corresponding protocol is transparently executed.
Each protocol has access to a representation of the system through
its metaobjects, by means of which it can also perform actions
on the corresponding ``real'' objects.

MOPs appear to constitute a promising technique for
a transparent, coherent, and effective adoption of some of the existing FT
mechanisms and techniques.
A number of studies confirm that MOPs
reach efficiency in some cases~\cite{KirB91,MMWY92}, though no experimental
or analytical evidence allows so far to estimate the practicality and
the generality of this approach: ``what reflective capabilities
are needed for what form of fault-tolerance, and to what extent these
capabilities can be provided in more-or-less conventional programming
languages, and allied to other structuring techniques [e.g. recovery blocks
or NVP] remain to be determined''~\cite{RaXu95}. In other words, it is still an open
question whether MOPs represent a practical solution
towards the effective integration of most of the existing fault
tolerance mechanisms in the user applications.

The above situation reminds the authors
of another one, regarding the ``quest'' for a novel computational paradigm
for parallel processing which would be capable of dealing effectively
with the widest class of problems, as the Von Neumann paradigm does
for sequential processing, though with the highest degree of efficiency
and the least amount of changes in the original (sequential) user code.
In that context, the concept of computational {\em grain\/} came up---some 
techniques were inherently looking at the problem ``with
coarse-grained glasses,'' i.e., at macroscopic level, others were considering
the problem exclusively at microscopical level.
One can conclude that MOPs offer an elegant system structure to
embed a set of non-functional services (including fault-tolerance
provisions) in an object-oriented program. 
It is still unclear whether this set is general enough
to host, efficaciously, many forms of fault-tolerance,
as is remarked for instance in~\cite{RaXu95,LiLo00}.
It is therefore difficult to establish a qualitative
assessment of attribute \SA{} for MOPs.

The run-time management of libraries of MOPs may be used
to reach satisfactory values for attribute \Aty. To the
best of the authors' knowledge,
this feature is not present in any language supporting MOPs.

As evident, the target application domain is the one
of object-oriented applications written with languages
extended with a MOP, such as Open \CPP{}.

As a final remark, we observe how the \textbf{cost} of MOP-compliant
fault-tolerant software design should include those related to the
acquisition of the extra competence and experience in MOP design
tools, reification, and custom programming languages.

\subsection{Enhancing or Developing Fault-Tolerance Languages}%
\label{Sect:Lang}
Another approach is given by working at the language level enhancing a
pre-existing programming language or developing an ad hoc distributed 
programming language so that it hosts specific 
fault-tolerance provisions. The following two sections cover these topics.

\subsubsection{Enhancing Pre-existing Programming Languages}%
\label{SubSect:EnPrLang}
Enhancing a pre-existing programming language means
augmenting the grammar of a wide-spread language such as C
or Pascal so that it directly supports features that can be used
to enhance the dependability of its programs, e.g., 
recovery blocks~\cite{Shr78}.

In the following, four classes of systems based on this approach are
presented: Arjuna, Sina, Linda, and FT-SR. 
All of them constitute provisions to develop distributed fault-tolerant systems.

\paragraph{The Arjuna Distributed Programming System}
Arjuna is an object-ori\-en\-ted system for portable distributed
programming in \CPP{}~\cite{Par90,Shr95}. It can be considered as
a clever blending of useful and widespread tools, techniques,
and ideas---as such, it is a good example of the evolutionary 
approach towards application-level software fault-tolerance.
It exploits remote procedure calls~\cite{BiNe84a}
and UNIX daemons. On each node of the system
an object server connects client objects to objects
supplying services. The object server also takes care of
spawning objects when they are not yet running (in this case
they are referred to as ``passive objects'').
Arjuna also exploits a ``naming service'', by means of which client
objects request a service ``by name''. This transparency
effectively supports object migration and replication.

As done in other systems, Arjuna makes use of stub generation to specify
remote procedure calls and remote manipulation of objects. A nice feature of 
this system is that the stubs are derived automatically from the \CPP{} header
files of the application, which avoids the need of a custom interface
description language.

Arjuna offers the programmer means for dealing
with atomic actions (via the two-phase commit protocol)
and persistent objects. The core class hierarchy of Arjuna appears 
to the programmer as follows~\cite{Par90}:
	StateManager
		LockManager
			User-Defined Classes
		Lock
			User-Defined Lock Classes
		AtomicAction
		AbstractRecord
			RecoveryRecord
			LockRecord
			and other management record types
	etc.

Unfortunately, it requires the programmers
to explicitly deal with tools to save and restore the state, to
manage locks, and to declare in their applications instances of the
class for managing atomic actions. As its authors state,
in many respects Arjuna asks the programmer to be aware of
several complexities\footnote{The Arjuna stub generator attempts
	to compensate for these problems are far as it can automatically
	but there are cases where assistance from the programmer is required. For example, heterogeneity
	is handled by converting all primitive types to a standard format understood by both
	caller and receiver.}---as such, it is prejudicial to transparency
and separation of design concerns. On the other hand, its good design
choices result in an effective, portable environment.


\paragraph{The SINA Extensions}
The SINA~\cite{AkDij91} object-oriented language implements the so-called
composition filters object model, a modular
extension to the object model. In SINA, each object is equipped with
a set of ``filters''. Messages sent to any object are trapped
by the filters of that object. These filters possibly manipulate the
message before passing it to the object. SINA is a language
for composing such filters---its authors refer to it
as a ``composition filter language''. It also supports
meta-level programming through the reification of messages.
The concept of composition filters
allows to implement several different ``behaviours'' corresponding
to different non-functional concerns.
SINA has been designed for being attached to existing languages:
its first implementation, SINA/st, was for Smalltalk.
It has been also implemented for \CPP{}~\cite{Gla95}---the extended
language has been called \CPP{}/CF. 
A preprocessor is used to translate a
\CPP{}/CF source into standard \CPP{} code.


\paragraph{Fault-Tolerant Linda Systems}\label{Sect:Linda}
The Linda~\cite{CaGe89a,CaGe89b} approach 
adopts a special model of communication,
known as generative communication~\cite{Ge85}.
According to this model, communication is still carried out
through messages, though
messages are not sent to one or more addressees, and eventually
read by these latter---on the contrary, messages are included in a
distributed (virtual) shared memory, called tuple space, where every
Linda process has equal read/write access rights.
A tuple space is some sort of a shared relational database for storing
and withdrawing special data objects called tuples, sent by
the Linda processes. Tuples are basically lists of objects identified
by their contents, cardinality and type. Two tuples match if they have
the same number of objects, if the objects are pairwise equal for what
concerns their types, and if the memory cells associated to the objects
are bitwise equal.
A Linda process inserts, reads, and withdraws tuples via blocking or
non-blocking primitives. Reads can be performed supplying a template
tuple---a prototype tuple consisting of constant fields and
of fields that can assume any value. A process trying to access a missing tuple
via a blocking primitive enters a wait state that continues until any
tuple matching its template tuple
is added to the tuple space. This allows processes to synchronise.
When more than one tuple matches a template, the choice of which
actual tuple to address is done in a non-deterministic way.
Concurrent execution of processes is supported through the concept
of ``live data structures'': tuples requiring the execution of one or
more functions can be evaluated on different processors---in a sense,
they become active, or ``alive''. Once the evaluation has finished, 
a (no more active, or passive) output tuple is entered in the tuple space.

Parallelism is implicit in Linda---there is no explicit notion of network,
number and location of the system processors, though Linda has been
successfully employed in many different hardware architectures and
many applicative domains, resulting in a powerful programming tool that
sometimes achieves
excellent speedups without affecting portability issues. Unfortunately
the model does not cover the possibility of failures---for instance, the
semantics of its primitives are not well defined in the case of a processor
crash, and no fault-tolerance means are part of the model. Moreover, in its
original form, Linda only offers single-op atomicity~\cite{BaSc95},
i.e., atomic execution for only a single tuple space operation. With
single-op atomicity it is not possible to solve problems arising
in two common Linda programming paradigms when faults occur: both
the distributed variable and the replicated-worker paradigms
can fail~\cite{BaSc95}. As a consequence,
a number of possible improvements have been investigated to support
fault-tolerant parallel programming in Linda. Apart from design choices
and development issues, many of them implement stability of the
tuple space (via replicated state machines~\cite{Sch90} kept consistent
via ordered atomic multicast~\cite{biscst})~\cite{BaSc95,XuLi89,PaTHR93},
while others aim at combining multiple tuple-space operations into 
atomic transactions~\cite{BaSc95,AnSh91,CaDu92}. Other techniques have
also been used, e.g., tuple space checkpoint-and-rollback~\cite{Kam91}.
The authors also proposed an augmented Linda
model for solving inconsistencies related to failures occurring in a
replicated-worker environment and an algorithm for
implementing a resilient replicated worker strategy for message-passing
farmer-worker applications. This algorithm can mask failures
affecting a proper subset of the set of workers~\cite{DeDL98g}.

Linda can be described as an extension that can be added to an
existing programming language. The greater part of these extensions requires
a preprocessor translating the extension in the host language.
This is the case, e.g., 
for FT-Linda~\cite{BaSc95}, PvmLinda~\cite{DeMS94}, C-Linda~\cite{Ber89},
and MOM~\cite{AnSh91}. A counterexample is, e.g., 
the POSYBL system~\cite{Sch91}, which implements Linda primitives
with remote procedure calls, and requires the user to supply
the ancillary information for distinguishing tuples.

\paragraph{FT-SR}
FT-SR~\cite{ScTh95} is basically an attempt to augment
the SR~\cite{SR} distributed programming language with mechanisms
to facilitate fault-tolerance. FT-SR is based on the concept
of fail-stop modules (FSM). A FSM is defined as an abstract unit of
encapsulation. It consists of a number of threads that export
a number of operations to other FSMs. The execution of operations
is atomic. FSM can be composed so to give rise to complex FSMs.
For instance it is possible to replicate a module $n>1$ times and set up
a complex FSM that can survive to $n-1$ failures. Whenever a failure
exhausts the redundancy of a FSM, be it a simple or complex FSM, a
failure notification is automatically sent to a number of other FSMs
so as to trigger proper recovery actions. This feature explains the name
of FSM: as in fail-stop processors, either the system is correct
or a notification is sent and the system stops its functions.
This means that the computing model of FT-SR guarantees, to some
extent, that in the absence of explicit failure notification,
commands can be assumed to have been processed correctly.
This greatly simplifies program development because it masks the
occurrence of faults, offers guarantees that no erroneous results are produced,
and encourages the design of complex, possibly dynamic failure
semantics (see Sect.~\ref{Sect:Paradox}) based on failure notifications.
Of course this strategy
is fully effective only under the hypothesis of perfect failure 
detection coverage---an assumption that sometimes may be found
to be false.
\paragraph{Conclusions}
\begin{figure}
\centerline{\includegraphics[width=12cm]{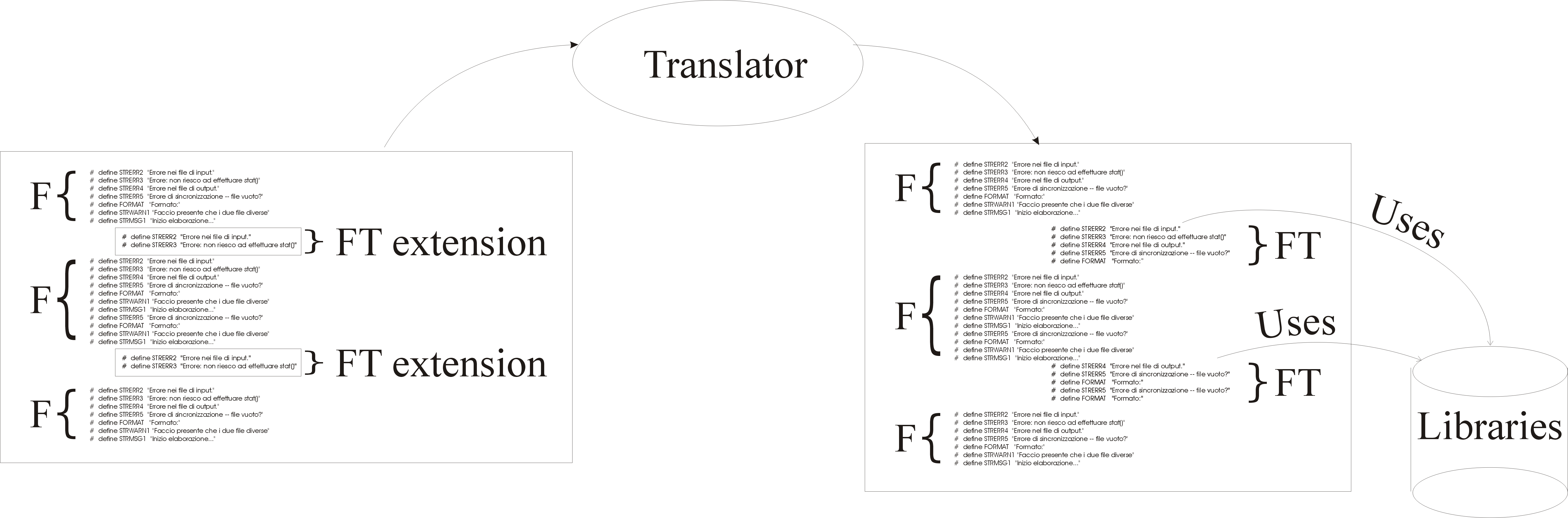}}
\caption{A fault-tolerant program according to the enhanced language approach. Note
how in this case a translator decomposes the program into a SV system.}\label{f:phi-EL}
\end{figure}
Figure~\ref{f:phi-EL} synthesizes the main characteristics of
the enhanced language approach: A compiler or, as in the picture, a translator,
produces a new fault-tolerant program. In the case in the picture the translated program
belongs to class SV (see Sect.~\ref{SubSect:SVSFT}). Note how few clearly identifiable
``FT'' extensions are translated into larger sections of
fault-tolerant code. As characteristics of an SV system, this fault-tolerant code
is indistinguishable from the functional code of the application.

We can conclude by stating that the approach of designing
fault-tolerance enhancements for a pre-existing programming
language does imply an explicit code intrusion: The 
extensions are designed with the explicit purpose to
host a number of fault-tolerance provisions within the
single source code. We observe, though, that being explicit, this code intrusion
is such that the fault-tolerant code is generally
easy to locate and distinguish from the functional
code of the application. Hence, attribute \SC{} may be
positively assessed for systems belonging to this category.
Following a similar reasoning and observing Fig.~\ref{f:phi-EL}
we can conclude that the design and
maintenance \textbf{costs} of this approach are in general less than
those characterising SV.

On the contrary, the problem of hosting an adequate structure
for ALFT can be complicated by the syntax constraints in the
hosting language.
This may prevent incorporation of a wide set
of fault-tolerance provisions
within a same syntactical structure. One can conclude that
in this case attribute \SA{} does not reach satisfactory values---at least
for the examples considered in this section.

Enhancing a pre-existing language is an
\emph{evolutionary\/} approach: in so doing, portability problems
are weakened---especially when the extended grammar is
translated into the plain grammar, e.g., via
a preprocessor---and can be characterised by good 
execution efficiency~\cite{And85,ScTh95}.

The approach is generally applicable, though the application
must be written (or rewritten) using the enhanced language.
Its adaptability (attribute \Aty) is in general unsatisfactory, because
at run-time the fault-tolerant code is indistinguishable
from the functional code.

As a final observation, we remark how the four cases that have
been dealt with in Sect.~\ref{SubSect:EnPrLang} all stem from the
domain of distributed/concurrent programming, which shows the
important link between fault tolerance issues and distributed
computing.


\subsubsection{Developing Novel Fault-Tolerance Programming Languages}\label{DL}
The adoption of a custom-made language especially conceived
to write fault-tolerant distributed software is discussed
in the rest of this subsection.
\paragraph{ARGUS}
Argus~\cite{Lis88} is a distributed object-oriented programming language and 
operating system. Argus was designed to support application programs such as
banking systems. To capture the object-oriented nature of such programs, 
it provides a special kind of objects,
called guardians, which perform user-definable actions in response
to remote requests.  To solve the problems of
concurrency and failures, Argus allows computations to run as
atomic transactions. Argus' target application domain is the one
of transaction processing.

\paragraph{The Correlate Language}
The Correlate object-oriented language~\cite{Rob99} adopts the concept of
an active object, defined as an object that has control over the
synchronisation of incoming requests from other objects. Objects are
active in the sense that they do not process their requests immediately---they
may decide to delay a request until it is accepted, i.e., until a given
precondition (a guard) is met---for instance, a mailbox object may refuse
a new message in its buffer until an entry becomes available in it.
The precondition is a function of the state of the object and the invocation
parameters---it does not imply interaction with other objects and has
no side effects. If a request cannot be served according to
an object's precondition, it is saved into a buffer until it
becomes serviceable, or until the object is destroyed. Conditions
such as an overflow in the request buffer are not dealt with in~\cite{Rob99}.
If more than a single request becomes serviceable by an object, the
choice is made non-deterministically. Correlate uses a
communication model called ``pattern-based group communication''---communication
goes from an ``advertising object'' to those objects that declare
their ``interest'' in the advertised subject. This is similar
to Linda's model of generative communication, introduced in 
Sect.~\ref{Sect:Linda}.
Objects in Correlate are autonomous, in the sense that they may not only react
to external stimuli but also give rise to autonomous operations motivated by
an internal ``goal''. When invoking a method, the programmer can choose
to block until the method is fully executed (this is called synchronous interaction),
or to execute it ``in the background'' (asynchronous interaction).
Correlate supports MOPs. It has been effectively used to
offer transparent support for transaction, replication, and checkpoint-and-rollback.
The first implementation of Correlate consists of a translator
to plain Java plus an execution environment, also written in Java.

\paragraph{Fault-Tolerance Attribute Grammars}\label{Sect:FTAG}
The system models for application-level software fault-tolerance
encountered so far all have their basis in an imperative language.
A different approach is based on the use of
functional languages. This choice translates into a program
structure that allows a straightforward inclusion of a means for
fault-tolerance, with high degrees of transparency
and flexibility. Functional models that appear particularly
interesting as system structures for software fault-tolerance
are those based on the concept of attribute grammars~\cite{Paa95}.
This paragraph briefly introduces the
model known as FTAG (fault-tolerant attribute grammars)~\cite{SKS96},
which offers the designer a large set of fault-tolerance mechanisms.
A noteworthy aspect of FTAG is that its authors explicitly address
the problem of providing a syntactical model for
the widest possible set of fault-tolerance provisions and paradigms,
developing coherent abstractions of those mechanisms while maintaining
the linguistic integrity of the adopted notation. in other words,
optimising the value of attribute \SA{} is one of the design goals
of FTAG.

FTAG regards a computation as a collection of pure mathematical functions
known as modules. Each module has a set of input values, called
inherited attributes, and of output variables, called synthesised attributes.
Modules may refer to other modules. When modules do not refer to any
other module, they can be performed immediately. Such modules are called
primitive modules. On the other hand, non-primitive modules
require other modules to be performed first---as a consequence, an FTAG
program is executed by decomposing a ``root'' module into its basic
submodules and then applying this decomposition process recursively
to each of the submodules. This process goes on until all primitive
modules are encountered and executed. The execution graph is clearly a tree
called computation tree. This approach presents many benefits,
e.g., as the order in which modules are decomposed is exclusively
determined by attribute dependencies among submodules, a computation
tree can be turned straightforwardly into a parallel process.

The linguistic structure of FTAG allows the integration of a number
of useful fault-tolerance features that address the whole range
of faults---design, physical, and interaction faults. 
One of these features is called redoing. Redoing replaces a portion
of the computation tree with a new computation. This is useful
for instance to eliminate the effects of a portion of the computation tree
that has generated an incorrect result, or whose executor has crashed.
It can be used to implement easily ``retry blocks'' and recovery blocks
by adding ancillary modules that test whether the original module behaved
consistently with its specification and, if not, give rise to a
``redoing'', a recursive call to the original module.

Another relevant feature of FTAG is its support for replication,
a concept that in FTAG translates into a decomposition of a module
into $N$ identical submodules implementing the function to replicate.
Such approach is known as replicated decomposition, while involved
submodules are called replicas. Replicas are executed according
to the usual rules of decomposition, though only one of the generated results
is used as the output of the original module. Depending on the
chosen fault-tolerance strategy, this output can be, e.g., the
first valid output or the output of a demultiplexing function, e.g.,
a voter. It is worth remarking that no syntactical changes are needed, only a
subtle extension of the interpretation so to allow the involved submodules
to have the same set of inherited attributes and to generate a collated
set of synthesised attributes.

FTAG stores its attributes in a stable object base or in primary memory
depending on their criticality---critical attributes can then be transparently
retrieved from the stable object base after a failure. Object versioning is
also used, a concept that facilitates the development of checkpoint-and-rollback strategies.

FTAG provides a unified linguistic structure that effectively supports
the development of fault-tolerant software. Conscious of the importance
of supporting the widest possible set of fault-tolerance means, its
authors report in the cited paper how they are investigating the inclusion
of other fault-tolerance features and trying to synthesise new
expressive syntactical structures for FTAG---thus further improving
attribute \SA.

Unfortunately, the widespread adoption of
this valuable tool is conditioned by the limited acceptance and spread
of the functional programming paradigm outside of academia.

\begin{figure}
\centerline{\includegraphics[width=9cm]{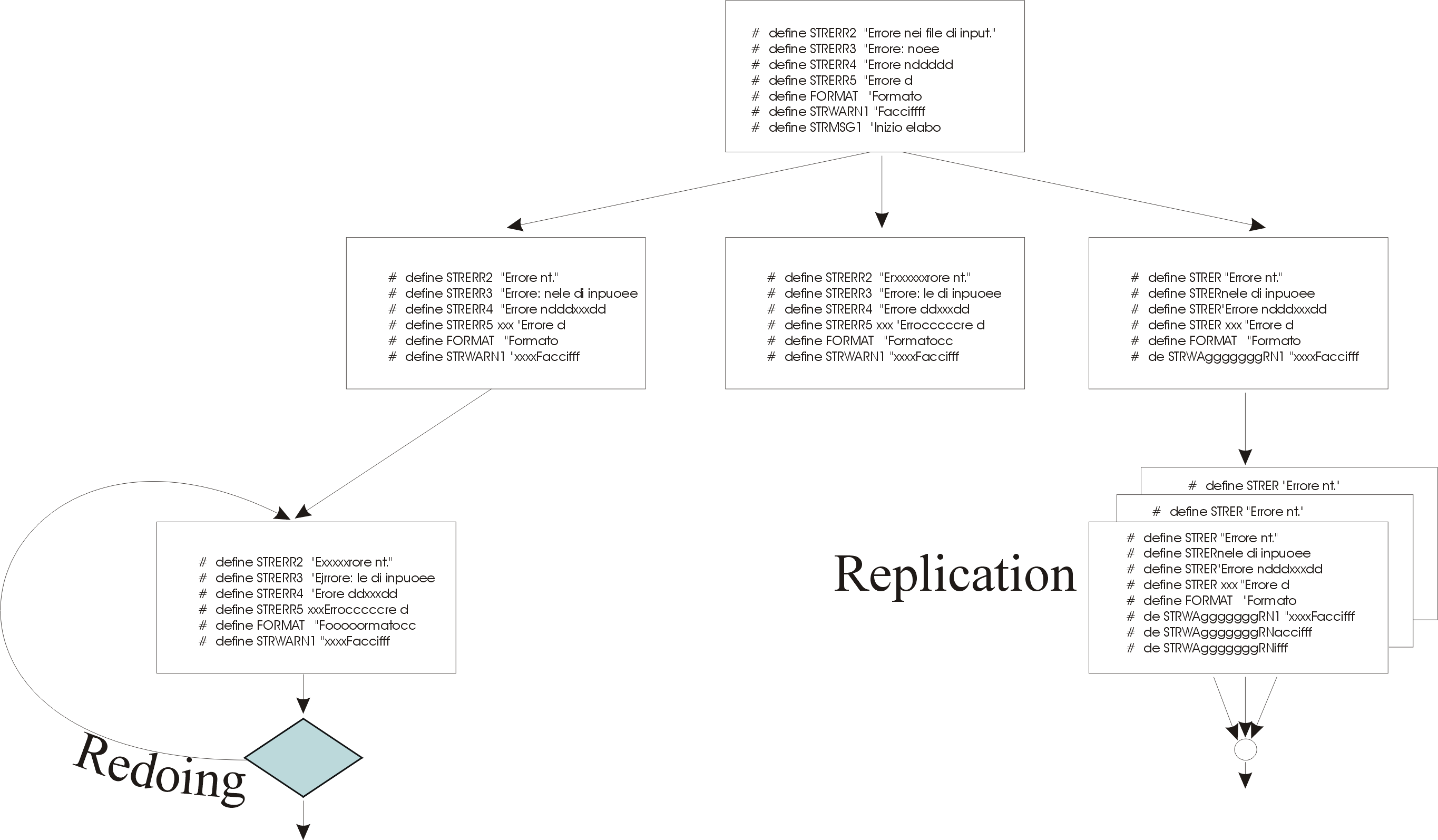}}
\caption{A fault-tolerant program according to a FTAG system.}\label{f:phi-ftag}
\end{figure}

\paragraph{Conclusions}
Synthesizing in a single meaningful picture the main characteristics of the 
approach that makes use of custom fault-tolerance languages is very difficult,
for the design freedom translates into entities that may have few points in
common. Figure~\ref{f:phi-ftag} for instance synthesizes the characteristics of
FTAG.

The ad hoc development of a fault-tolerance programming language
allows
optimal values of attribute \SA{}
to be reached in some cases.
The explicit, controlled intrusion of fault-tolerant code 
explicitly encourages
the adoption of high-level fault-tolerance provisions
and requires dependability-aware design processes,
which translates into a positive assessment for attribute \SC.
On the contrary, with the same reasoning of Sect.~\ref{SubSect:EnPrLang},
attribute \Aty{} can be in general assessed as unsatisfactory\footnote{%
	In the case of FTAG, though, one could design
	a run-time interpreter that dynamically ``decides'' the
	``best'' values for the parameters of the fault-tolerance
	provisions being executed---where ``best'' refers to the
	current environmental conditions.}.

The target application domain for this approach is restricted
by the characteristics of the hosting language and of its
programming model. Obviously this also requires the application
to be (re-)written using the hosting language. The acquisition of
know-how in novel design paradigms and languages is likely to
have an impact on development \textbf{costs}.

\subsection{Aspect-oriented Programming Languages}\label{Sect:AOP}
Aspect-oriented programming (AOP)~\cite{KLM97} is a 
programming methodology and a structuring technique
that explicitly addresses,
at a system-wide level, the problem of the best code structure
to express different, possibly conflicting design goals
such as high performance, optimal memory usage, and 
dependability.

Indeed, when coding a non-functional service within
an application---for instance a system-wide error handling
protocol---using either a procedural or an object-oriented
programming language, one is required to decompose the
original goal, in this case a certain degree of dependability,
into a multiplicity of fragments scattered among a
number of procedures or objects. This happens because
those programming languages only provide abstraction and
composition mechanisms to cleanly support the
functional concerns. In other words, specific
non-functional goals, such as high performance, cannot be
easily captured into a single unit of functionality
among those offered by a procedural or object-oriented
language, and must be fragmented and
intruded into the available units of
functionality. As already observed, this code intrusion
is detrimental to maintainability and portability of
both functional and non-functional services (the latter
called ``aspects'' in AOP terms). These aspects tend to crosscut
the system's class and module structure rather than
staying, well localised, within one of these
unit of functionality, e.g., a class. This increases
the complexity of the resulting systems.

The main idea of AOP is to use:
\begin{enumerate}
\item A ``conventional'' language
  (that is, a procedural, object-oriented, or functional
  programming language) to code the basic functionality.
  The resulting program is called component program.
  The program's basic functional units are called
  components.
\item A so-called
  aspect-oriented language to implement
  given aspects by defining specific interconnections
  (``aspect programs'' in AOP lingo) among the components
  in order to address various systemic concerns.
\item An aspect weaver, that takes as input
  both the aspect and the component programs and
  produces with those (``weaves'') an output program 
  (``tangled code'') that addresses specific aspects.
\end{enumerate}
The weaver first generates a data flow graph from the
component program. In this graph, nodes represent
components, and edges represent data flowing
from one component to another. Next, it executes
the aspect programs. These programs edit the
graph according to specific goals, collapsing
nodes together and adjusting the corresponding
code accordingly. Finally, a code generator
takes the graph resulting from the previous
step as its input and translates it into
an actual software package written, e.g.,
for a procedural language such as C.
This package is only meant to be compiled
and produce the ultimate executable code
fulfilling a specific aspect such as, e.g.,
higher dependability.

In a sense, AOP systematically automatises and
supports the process to adapt an existing code
so that it fulfils specific aspects.
AOP may be defined as a software engineering
methodology supporting those adaptations
in such a way that they
do not destroy the original design and
do not increase complexity.
The original idea of AOP is a clever
blending and generalisation of the ideas
that are at the basis, for instance,
of optimising compilers, program transformation
systems, MOPs, and of literate
programming~\cite{Knuth92}.

\subsubsection{AspectJ} AspectJ is probably the very
first example of aspect-oriented
language~\cite{Kic00,LiLo00}. Developed as a Xerox PARC
project, AspectJ can be defined as an aspect-oriented
extension to the Java programming language.
AspectJ provides its users with the concept of a ``join points'',
i.e., relevant points in a program's dynamic call graph.
Join points are those that mark the code regions that
can be manipulated by an aspect weaver (see above).
In AspectJ, these points can be
\begin{itemize}
\item method executions,
\item constructor calls,
\item constructor executions,
\item field accesses, and
\item exception handlers.
\end{itemize}
Another extension to Java is AspectJ's support of the
Design by Contract methodology~\cite{Mey97}, where
contracts~\cite{Hoa69} define a set of pre-conditions,
post-conditions, and invariants, that
determine how to use and
what to expect from a computational entity.

A study has been carried out on the
capability of AspectJ as an AOP language supporting
exception detection and handling~\cite{LiLo00}. It has been
shown how AspectJ can be used to develop so-called
``plug-and-play'' exception handlers: libraries of
exception handlers that can be plugged into many
different applications. This translates into better
support for managing different configurations
at compile-time. This addresses one of the
aspects of attribute \Aty{} defined in Sect.~\ref{Sect:StrAttr}.

\subsubsection{AspectWerkz} Recently a new stream of research activity has
been devoted to dynamic AOP. An interesting example of this trend is
AspectWerkz~\cite{BoVa04,Vas04}, defined by its authors as
``a dynamic, lightweight and high-performant AOP framework for
Java''~\cite{Bon04}. 
AspectWerkz utilizes bytecode modification to weave classes at project 
build-time, class load time or runtime. This capability means that
the actual semantics of an AspectWerkz code may vary dynamically
over time, e.g., as a response to environmental changes. This translates
into good support towards \Aty.

Recently the AspectJ and AspectWerkz projects have agreed to work 
together as one team to produce a single aspect-oriented
programming platform building on their complementary strengths 
and expertise.

\subsubsection{AspectC++}
A recent project, AspectC++~\cite{SpLU05,AspectCpp}, proposes an aspect-oriented implementation of
C++ which appears to achieve most of the positive properties of
the other Java-based approaches and adds to this efficiency and
good performance.

\begin{figure}
\centerline{\includegraphics[width=12cm]{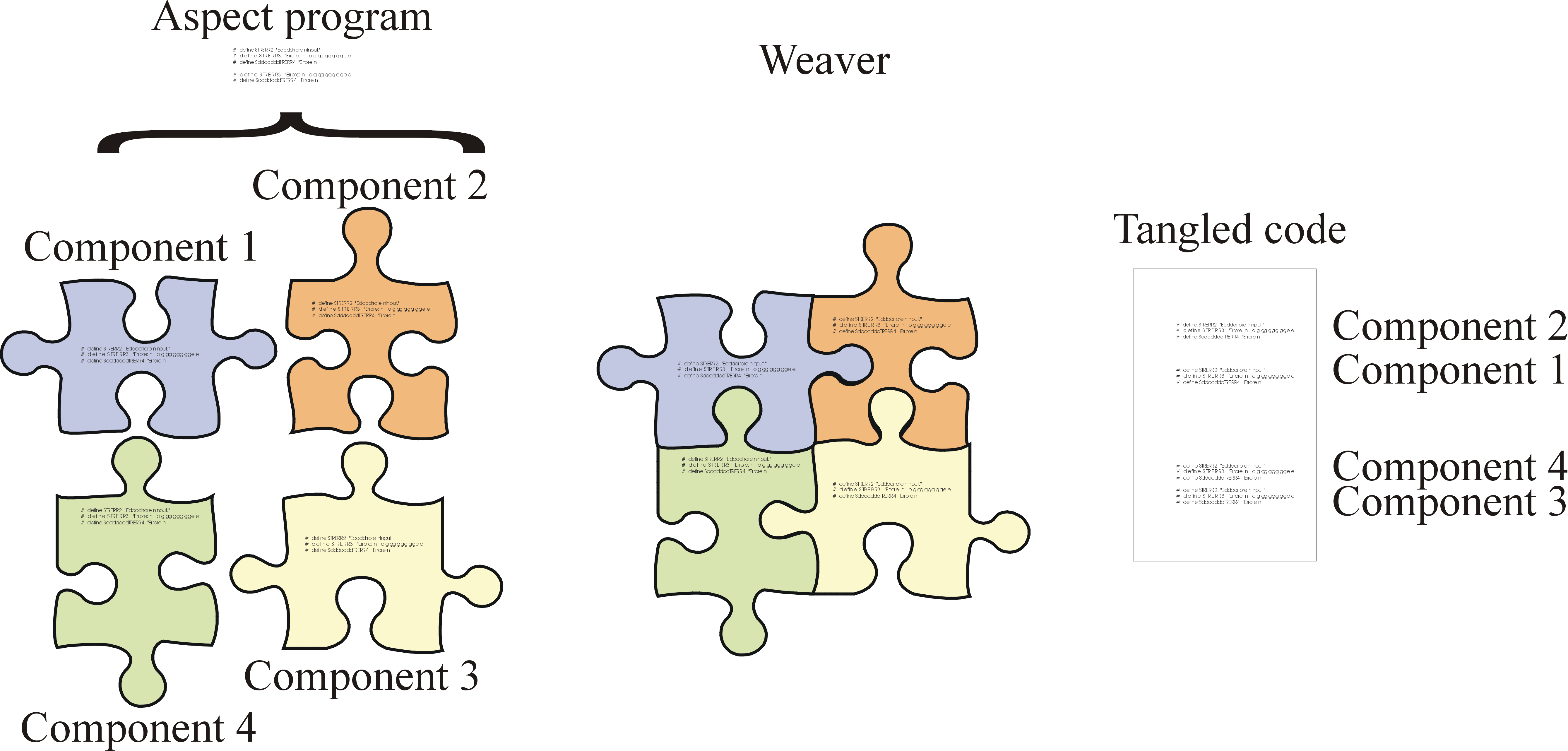}}
\caption{A fault-tolerant program according to an AOP system.}\label{f:phi-aop}
\end{figure}

\subsubsection{Conclusions}
Figure~\ref{f:phi-aop}
synthesizes the main characteristics of AOP: it allows to decompose, select, and assemble
components according to different design goals. This has been represented by drawing
the components as pieces of a jigsaw puzzle created by the aspect program and assembled by
the weaver into the actual source code. AOP addresses explicitly code re-engineering,
which in principle should allow to reduce considerably \textbf{maintenance costs}.

AOP is a relatively recent approach to 
software development. AOP can in principle
address any application domain and can use
a procedural, functional or object-oriented
programming language as component language.
The isolation and coding of aspects requires
extra work and expertise that may be well
payed back by the capability of addressing
new aspects while keeping a single unmodified
and general design. This said, we must remark how
some researchers questioned the adequacy of AOP
as an effective paradigm for handling failure---at least
for the domain of transaction processing~\cite{KiGu02}.

For the time being it is
not yet possible to tell whether AOP will
spread out as a programming paradigm among
academia and industry the way object-oriented
programming has done since the Eighties.
The many qualities of AOP are currently
being quantitatively assessed, both with
theoretical studies and with practical
experience, and results seem encouraging.
Furthermore, evidence of an increasing interest in AOP
is given by the large number of research papers and
conferences devoted to this interesting subject.

From the point of view of the dependability
aspect, one can observe that AOP exhibits
optimal \SC{} (``by construction'', in a sense~\cite{KiMe05}),
and that recent results show that
attribute \Aty{} can in principle reach good values
when making use of run-time
weaving~\cite{Vas04}, often realized by dynamic bytecode manipulation.
The work by Ostermann~\shortcite{Ost} is an interesting survey
on this subject.

The adequacy at fulfilling attribute
\SA{} is indeed debatable also because, to date,
no fault-tolerance aspect languages have been
devised\footnote{For instance, AspectJ only addresses exception error detection
   and handling. Remarkably enough, the authors of
   a study on AspectJ and its support to this field
   conclude~\cite{LiLo00} that ``whether the properties of
   AspectJ [documented in this paper] lead to programs
   with fewer implementation errors and that can be changed easier,
   is still an open research topic that will require serious
   usability studies as AOP matures''.}---which may possibly
be an interesting research domain. 


\subsection{The Recovery Meta-Program}\label{Sect:RMP}
\begin{figure}
\centerline{\includegraphics[width=11cm]{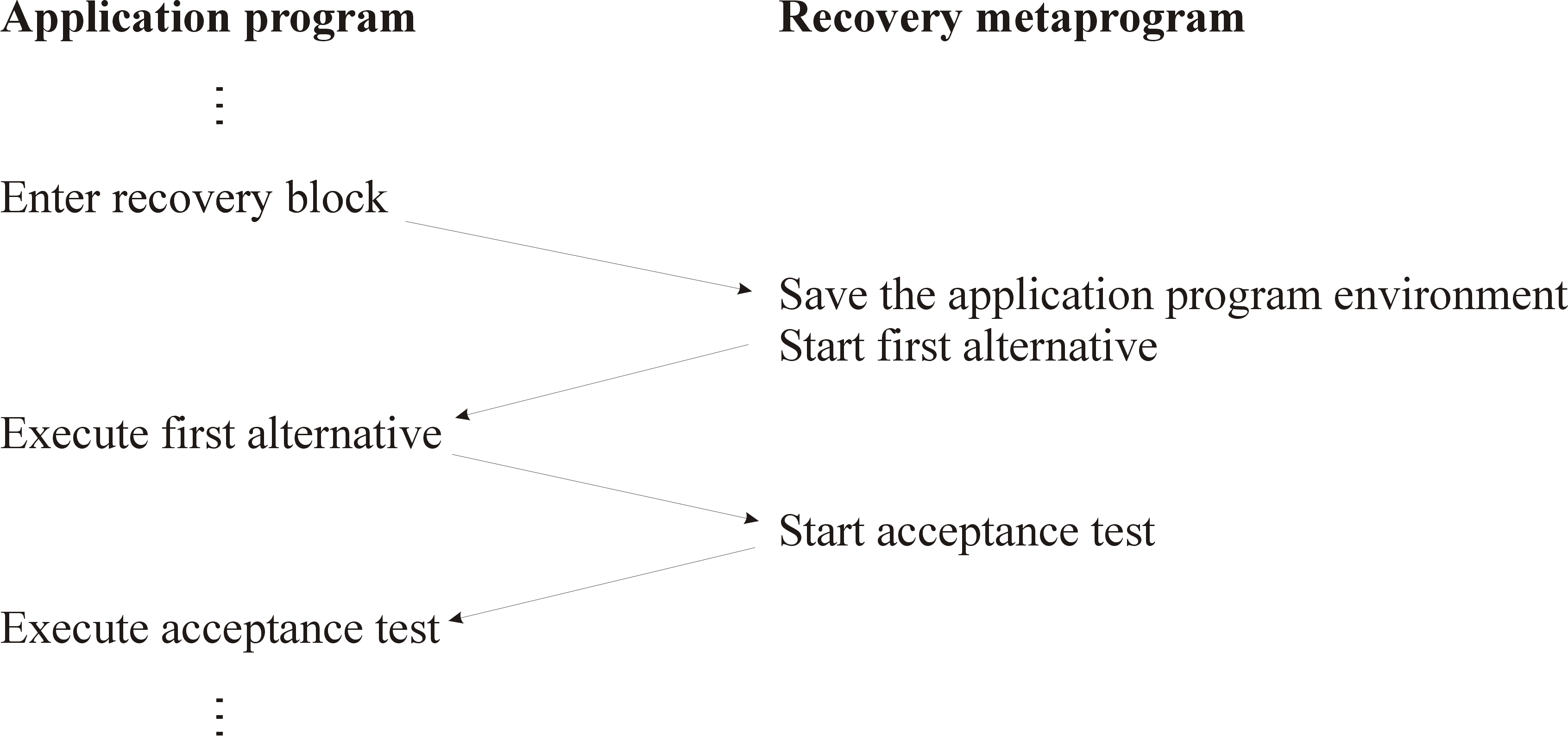}}
\caption{Control flow between the application program and RMP while 
executing a fault-tolerance strategy based on recovery blocks.}
\label{Fig:RMP}
\end{figure}

The Recovery Meta-Program (RMP)~\cite{AnDo90} is a mechanism
that alternates
the execution of two cooperating processing contexts. The concept behind its
architecture can be captured by means of the idea of a debugger, or a
monitor, which:
\begin{itemize}
 \item is scheduled when the application is stopped at some
       breakpoints,
 \item executes some program, written in a specific
       language,
 \item and finally returns the control to the application context,
       until the next breakpoint is encountered.
\end{itemize}
Breakpoints outline portions of code relevant to specific fault-tolerance
strategies---for instance, breakpoints can be used to specify alternate blocks
or acceptance tests of recovery blocks (see 
Sect.~\ref{Sect:RecoveryBlock})---while programs are implementations of
those strategies, e.g., of recovery blocks or $N$-version programming.
The main benefit of RMP is in the fact that, while breakpoints require a
(minimal) intervention of the functional-concerned programmer, RMP scripts
can be designed and implemented without the intervention and even the
awareness of the developer.
In other words, RMP guarantees a good separation of design concerns.
As an example, Fig.~\ref{Fig:RMP} shows how recovery blocks can be
implemented in RMP:
\begin{itemize}
\item When the system encounters a breakpoint corresponding to
      the entrance of a recovery block, control flows to the RMP,
      which saves the application program environment and starts
      the first alternate.
\item The execution of the first alternate goes on until its end,
      marked by another breakpoint. The latter returns the control
      to RMP, this time in order to execute the acceptance test.
\item Should the test succeed, the recovery block is exited,
      otherwise control goes to the second alternate, and so forth.
\end{itemize}
Note how the fault-tolerance development \textbf{costs} here are basically those
for specifying the alternates and acceptance tests, while the remaining complexity
is charged to the RMP architecture entirely.

In RMP, the language to express the meta-programs is Hoare's
Communicating Sequential Processes language~\cite{Hoa78} (CSP).

\begin{figure}
\centerline{\includegraphics[width=7cm]{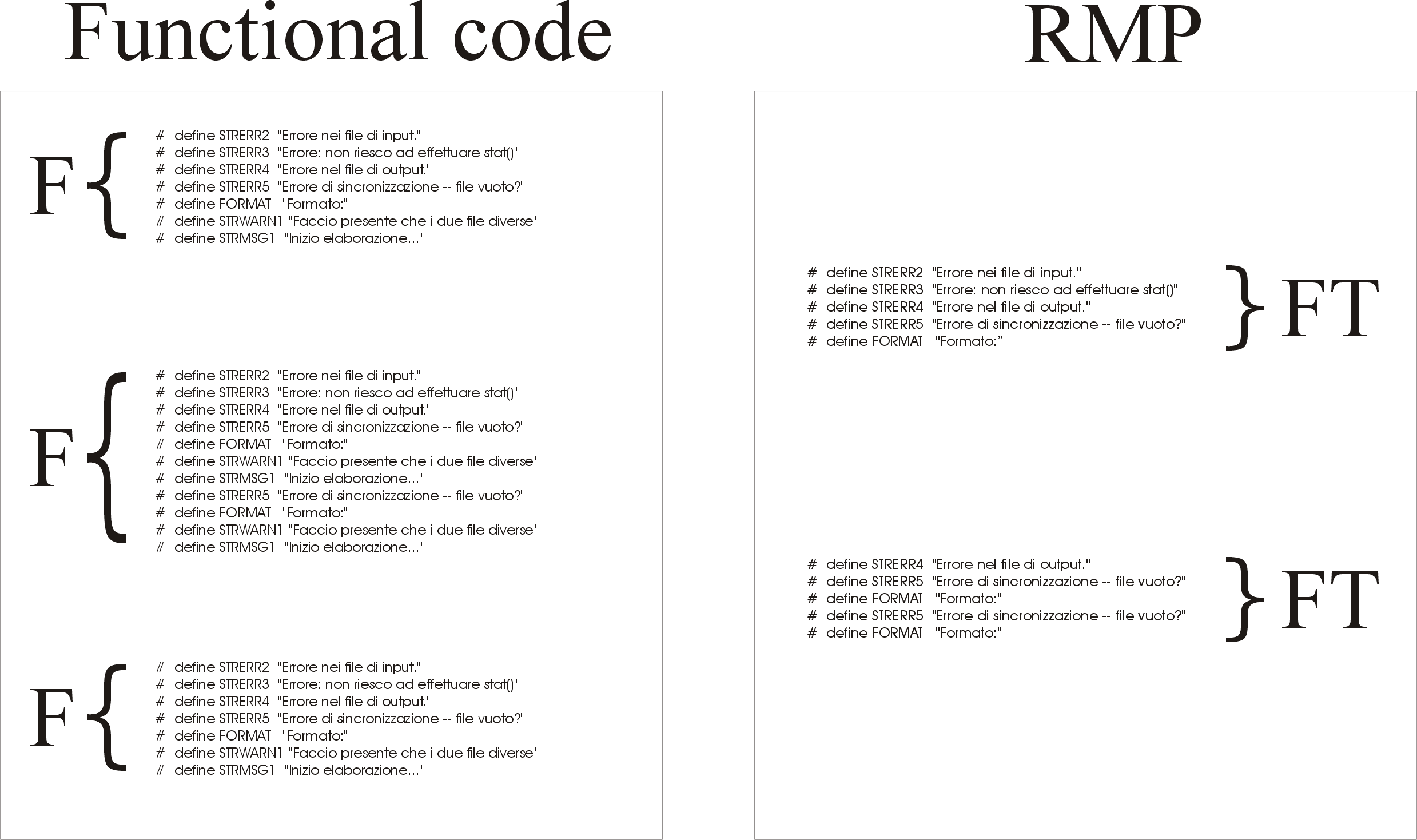}}
\caption{A fault-tolerant program according to the RMP system.}\label{f:phi-rmp}
\end{figure}
\subsection{Conclusions}
Figure~\ref{f:phi-EL} synthesizes the main characteristics of RMP: The fault-tolerance
code is in this case both logically and physically distinct from the functional code,
which means that the coding complexity and \textbf{costs} are considerably reduced.

In the RMP approach, all the technicalities related to the
management of the fault-tolerance provisions are coded in a
separate programming context. Even the language to code
the provisions may be different from the one used to express
the functional aspects of the application.
One can conclude that RMP is characterised by optimal \SC.

The design choice of using CSP to code the meta-programs
influences attribute \SA{} negatively.
Choosing a pre-existent formalism clearly presents many practical
advantages, though it means adopting a fixed, immutable syntactical
structure to express the fault-tolerance strategies.
The choice of a pre-existing general-purpose distributed
programming language such as CSP is therefore questionable, as it
appears to be rather difficult or at least cumbersome to use it
to express at least some of the fault-tolerance provisions.
For instance,
RMP proves to be an effective linguistic structure to express
strategies such as recovery blocks and
$N$-version programming, where the main components are coarse grain
processes to be arranged into complex fault-tolerance structures.
Because of the choice of a pre-existing language such as CSP,
RMP appears not to be the best choice for
representing provisions such as, e.g., atomic actions~\cite{JaCa85}.
This translates into very limited \SA{}.

Our conjecture is that the coexistence of two separate layers\label{conj1}
for the functional and the non-functional aspects could have been
better exploited to reach the best of the two approaches: using a
widespread programming language such as, e.g., C, for expressing
the functional aspect, while devising a custom language for 
dealing with non-functional requirements, e.g., a language
especially designed to express error recovery strategies. 

Satisfactory values for attribute \Aty{} cannot be
reached with the only RMP system developed so far,
because it does not foresee any dynamic management of the executable code.
Nevertheless it is definitely possible to design systems in which,
e.g., the recovery metaprogram changes dynamically so as to compensate
changes in the environment or other changes. This is the strategy
used in~\cite{DeBl05a} to set up a system structure for
adaptive mobile applications. Hence we chose to evaluate as
positive the value of \Aty{} for the RMP \emph{approach}.

RMP appears to be characterised by a large overhead
due to frequent context switching between the main application and
the recovery metaprogram~\cite{RaXu95}.
Run-time requirements may be jeopardised by these large overheads,
especially when it is difficult to establish time bounds for
their extent.
No other restrictions appear to be posed by RMP on the target
application domain.

\section{Conclusions}\label{Chap:end}
\begin{table}
\begin{center}
\begin{tabular}{|l|l|l|l|l|}
\hline
\textbf{Section}&\textbf{Approach}    &   \SC   &    \SA    &    \Aty\\ \hline
3.1.1           &SV                   &    poor    &     very limited     &     poor  \\
3.1.2           &MV (recovery blocks) &    poor    &     poor     &     poor  \\
3.1.2           &MV (NVP)             &    good    &     poor     &     poor  \\ \hline
3.2             &MOP                  &    optimal    &     positive?    &     positive  \\ \hline
3.3.1           &EL                   &    positive    &     poor     &     poor  \\
3.3.2           &DL                   &    positive    &     optimal     &     poor  \\ \hline
3.4             &AOP                  &    optimal    &     positive?    &     good  \\ \hline
3.5             &RMP                  &    optimal    &     very limited     &     positive  \\ \hline
\end{tabular}
\end{center}
\caption{A summary of the
qualitative assessments proposed in
Sect.~\ref{Chap:DesignTools}.
MV has been differentiated into recovery blocks (RB) and NVP.
EL is the approach of Sect.~\ref{SubSect:EnPrLang},
while DL is that of Sect.~\ref{DL}.}
\label{Tab:summary3}
\end{table}
Five classes of system structures for ALFT have been
described and critically reviewed, qualitatively, with
respect to the structural attributes \SC, \SA, and \Aty.
Table~\ref{Tab:summary3}
summarises the results of this survey
providing a comparison of the various approaches.
As can be seen from those summaries, no single approach exists today
that provide an optimal solution to the problems cumulatively referred to as
the system structure for application-level fault-tolerance.
We are currently working towards the definition of new models for
application-level fault-tolerance reaching high values of the three
attributes. A prototypic system is described in~\cite{DB07d}.

This paper has also highlighted the positive and negative aspects
of the evolutionary solution---using
a pre-existing language---with respect to a ``revolutionary''
approach---based on devising a custom made, ad hoc language.
As a consequence of these observations, it has been conjectured how
using an approach
based on two languages---one covering the functional
concerns and the other covering the fault-tolerance
concerns---it may be possible to address, within one\label{conj3}
efficacious linguistic structure, the widest set of
fault-tolerance provisions, thus providing optimal values for
the three structural attributes. This conjecture is currently
under verification~\cite{DeBl05a} in the framework of 
European project ARFLEX (``Adaptive Robots for Flexible Manufacturing Systems'')
and IBBT project QoE (``End-to-end Quality of Experience'').

\begin{ack}
We would like to express our gratitude to the Editor for the many insightful remarks and suggestions.
\end{ack}

\bibliographystyle{acmtrans}

\begin{received}
Received September 2003; revised August 2005; accepted August 2007.
\end{received}
\end{document}